\documentclass{IEEEtran}
\pdfoutput=1

\usepackage[usenames,dvipsnames]{color}
\usepackage{xspace}
\usepackage{multirow}
\usepackage{rotating}
\usepackage[table]{xcolor}

\usepackage{amsthm}
\usepackage{amsmath}
\usepackage{setspace}
\usepackage[font={footnotesize,bf},textformat=period,format=hang]{caption}
\usepackage[font=footnotesize,textformat=simple]{subcaption}
\usepackage{hhline}
\usepackage{multirow}
\usepackage[usenames,dvipsnames]{color}
\usepackage{graphicx}
\usepackage{cite}
\usepackage{url}
\usepackage{xspace}
\usepackage{paralist}
\usepackage{tikz}
\usepackage{pgfplots}
\usepackage{pgfplotstable}
\usepackage{etoolbox}

\providecommand\subparagraph{}
\usepackage{titlesec}

\usepackage{pifont}
\newcommand{\cmark}{\ding{51}}
\newcommand{\xmark}{\ding{55}}

\usetikzlibrary{patterns,arrows,positioning,shapes,calc}
\pgfplotsset{compat=newest, legend style={font=\scriptsize}}

\newcommand{\hide}[1] {}

\newtheoremstyle{exampstyle}
  {0.0\topsep} 
  {0.0\topsep} 
  {\itshape} 
  {} 
  {\bfseries} 
  {.} 
  {.5em} 
  {} 

\newtheorem{DEF}{Definition}[section]
\newtheorem{THEOREM}{Theorem}[section]
\newtheorem{LEMMA}[THEOREM]{Lemma}
\newtheorem*{remark}{Remark}

\definecolor{light-gray}{gray}{0.65}
\definecolor{dark-green}{HTML}{007F00}
\definecolor{mycyan}{HTML}{007FAA}

\newif\ifshowdata
\showdatatrue

\newif\ifshownotes

\ifshownotes

\newcommand{\mnote}[1] {{$\langle${\textcolor{red}{Mayur: \textbf{#1}}}$\rangle$}}
\newcommand{\knote}[1] {{$\langle${\textcolor{blue}{K.K.: \textbf{#1}}}$\rangle$}}
\newcommand{\jnote}[1] {{$\langle${\textcolor{dark-green}{Jia: \textbf{#1}}}$\rangle$}}
\newcommand{\mjnote}[1] {{$\langle${\textcolor{cyan}{M\&J: \textbf{#1}}}$\rangle$}}
\newcommand{\fnote}[1] {{$\langle${\textcolor{yellow}{Fu: \textbf{#1}}}$\rangle$}}
\newcommand{\snote}[1] {{$\langle${\textcolor{green}{S: \textbf{#1}}}$\rangle$}}

\else

\newcommand{\mnote}[1] {}
\newcommand{\knote}[1] {}
\newcommand{\jnote}[1] {}
\newcommand{\mjnote}[1] {}
\newcommand{\fnote}[1] {}
\newcommand{\snote}[1] {}
\fi

\newcommand{\vs}{\emph{vs.}\xspace}

\newcommand{\ie}{\emph{i.e.}\xspace}
\newcommand{\eg}{\emph{e.g.}\xspace}
\newcommand{\Eg}{\emph{E.g.}\xspace}
\newcommand{\etc}{\emph{etc.}\xspace}
\newcommand{\etal}{\emph{et al.}\xspace}

\newcommand{\first}{$1^{st}$\xspace}

\hide{
\makeatletter
\renewenvironment{itemize}
{\list{$\bullet$}{\leftmargin\z@ \labelwidth\z@
\itemindent-\leftmargin
}}
{\endlist} \makeatother
}

\hide{
\makeatletter
\renewenvironment{itemize} {
\begin{list}{$\bullet$}
    {
    \setlength{\itemsep}{0pt}
     \setlength{\parsep}{0pt}
     \setlength{\topsep}{0pt}
     \setlength{\partopsep}{0pt}
     \setlength{\leftmargin}{0.9em}
     \setlength{\labelwidth}{0.6em}
     \setlength{\labelsep}{0.4em}
    }
}
{\end{list}}
\newcounter{enumerates}
\renewenvironment{enumerate} {
\begin{list}{\arabic{enumerates}.}
    {\usecounter{enumerates}
    \setlength{\itemsep}{0pt}
     \setlength{\parsep}{0pt}
     \setlength{\topsep}{0pt}
     \setlength{\partopsep}{0pt}
     \setlength{\leftmargin}{1.1em}
     \setlength{\labelwidth}{0.9em}
     \setlength{\labelsep}{0.3em}
    }
}
{\end{list}}
\makeatother
}

\newcommand{\name}{SAID\xspace}
\newcommand{\conference}{arXiv\xspace}
\newcommand{\pt}{stall time\xspace}
\newcommand{\Pt}{Stall time\xspace}
\newcommand{\delivery}{ANP Delivery\xspace}

\title{\name: A Control Protocol for Scalable and Adaptive Information Dissemination in ICN}

\author{
\IEEEauthorblockN{Jiachen Chen\IEEEauthorrefmark{1}, Mayutan Arumaithurai\IEEEauthorrefmark{1}, Xiaoming Fu\IEEEauthorrefmark{1}, and K.K. Ramakrishnan\IEEEauthorrefmark{2}}\\
\IEEEauthorblockA{\IEEEauthorrefmark{1}University of G\"ottingen, Germany. Email: \{jiachen,arumaithurai,fu\}@cs.uni-goettingen.de}\\
\IEEEauthorblockA{\IEEEauthorrefmark{2}University of California, Riverside, USA. Email: kk@cs.ucr.edu}
}

\begin{document}
\maketitle
\thispagestyle{plain}
\pagestyle{plain}

\begin{abstract}
Information dissemination applications (video, news, social media, \etc)
with large number of receivers need to be efficient but also
have limited loss tolerance.
The new Information-Centric Networks (ICN) paradigm offers an alternative approach for
reliably delivering data by naming content and exploiting
data available at any intermediate point (\eg, caches). However, receivers are often heterogeneous, with widely
varying receive rates.
When using existing ICN congestion control mechanisms
with in-sequence delivery, a particularly thorny problem of receivers going \emph{out-of-sync} results in inefficiency and unfairness with heterogeneous receivers.
We argue that separating reliability from congestion control leads to more scalable, efficient and fair data dissemination, and propose \name, a Control Protocol for Scalable and Adaptive Information Dissemination in ICN.
To maximize the amount of data transmitted at the first attempt, receivers request \emph{any next packet} (ANP) of a flow instead of next-in-sequence packet,
independent of the provider's transmit rate.
This allows providers to transmit at an application-efficient rate, without
being limited by the slower receivers.
\name ensures reliable delivery to all receivers eventually, by
cooperative repair,
while preserving privacy without
unduly trusting other receivers.

\end{abstract}

\section{Introduction}
\label{sec-intro}
\hide{
- ip multicast in controlled environment
- why is receiver driven better
- Fig 4: highlight that it is cache size (allocated) for the flow
- should we mention: regardless of the cache replacement strategy
- 3D: is it clear which part is out of sync and which is not?
- tolerated out-of-sync?
- netowrk friendliness: Each branch should be able to receive its fair share. Should not consume more, but at the same time, not less too.
- should we say reliability is achieved for those applications that need it....
- Figure 7: make the states more clear
- show the ideal S1 and S2 bandwidth, keep y axis scale of SAID, ICP the same
- Figure 13: Mention the rocket fuel topology, no need to show the figure
- explain results with fundamental design choices that made us get it
- figures are too small
- change ICDCS to CONEXT, fonts exceeding space limit
- Fig 3: Doesn't have any mention about C51
}
Large scale information dissemination applications like video streaming (YouTube, NetfFlix, \etc), online social networks (Facebook, Twitter, \etc) and news/entertainment (CNN, BBC, RSS feeds, \etc) have become common. 
Many of these applications have limited loss tolerance and depend on the network to provide efficient, fair and reliable content distribution. 
While IP multicast was designed for large-scale information dissemination, the inability to have an effective congestion control solution, especially in the presence of heterogeneous capacity to receivers, has been a
limitation. 
The advent of Information-Centric Networks (ICN) offers us an opportunity to take a fresh look at the potential solution approaches. 

A key goal for publishers sending data at an application-efficient rate across the entire receiver population is to be not limited by the slower receivers.
Due to the absence of a network layer mechanism to control the delivery rate at the receiver end, previous solutions have either sought to push all the data onto the path, overlooking congestion and unfairness and using end-end unicast recovery, or seek to slow down the sending rate to the slowest receiver~\cite{pgmcc}. Alternatively, the use of unicast, with the associated inefficiencies, has become the norm.
The use of multicast at the application layer (\eg, SCRIBE~\cite{scribe}) exploits TCP's congestion control mechanisms for ensuring loss-free delivery on an end-end basis.
However, application layer solutions are network topology unaware and achieve lower efficiency (caused by the end-hosts replicating the packets rather than the routers)
 than what an effective multicast solution
could be expected to achieve.

Information-Centric Networking (ICN) is a new paradigm with the potential to achieve efficient large scale information delivery.
Content-Centric Networks (CCN~\cite{CCN}) or Named Data Networking (NDN~\cite{NDN})\footnote{In this work, we do not distinguish between NDN and CCN 
since we only refer to the fundamental communication model.} is a representative ICN approach.
The shift from a ``location-focused'' network to a ``content-centric'' network allows more efficient data sharing, thanks to name-based forwarding.
With in-network caches, NDN can exploit temporal locality among the consumers 
to outperform existing IP network-based approaches for a variety of information delivery situations.
Moreover, since entities in the network exchange information primarily on content names, the identities of the consumers need not be revealed, thus preserving user privacy.
Additionally, data integrity and provenance can be established based on the per-packet data generator-signature required in NDN.

Although NDN does not mandate a congestion control mechanism, most of the proposed solutions~\cite{ICP,hrtcp,ictp,contug,cctcp} choose to use a TCP-like receiver-driven mechanism to limit the number of requests (window) outstanding from a receiver.
Since the network maintains flow balance, where one Interest retrieves at most one Data packet, these mechanisms adapt the window using the Additive Increase Multiplicative Decrease (AIMD) principle, much like TCP~\cite{jacobson1988congestion, DECbit}.
Compared to a sender-driven rate control approach, such a receiver-driven approach has the benefit of the consumer being able to control the receive rate.
When considering efficient large scale data dissemination where every piece of data is consumed by a large number of receivers, TCP-like mechanisms for receiver-driven multicast can have significant shortcomings, especially with the path to the receivers having different capacities.
With such heterogeneity, a problem we observe is that of receivers being \textbf{out-of-sync}
\emph{even with optimal policies for managing the cache}.
The out-of-sync problem can be briefly described as follows: NDN routers cache content as they are forwarded. When there is temporal locality
of requests from different receivers, a router that has the cached content
can respond. 
However, with receiver heterogeneity,
the requests from receivers even for the same data items can diverge over time. Requests from the faster receivers can be well
ahead of those from slower receivers.
Eventually, when the gap between the faster and slower receivers becomes too large,
their requests can no longer be aggregated at the intermediate routers or satisfied by the cache.
The slower receivers' requests will have to be satisfied by the content provider (via retransmissions) as a separate flow.
These retransmissions will compete for the bottleneck bandwidth and the overall throughput can therefore dramatically reduce.
This is a fundamental issue as long as there are heterogeneous receivers and routers with limited cache sizes. The problem can be exacerbated with scale, thus occurring
even more often in the core of the network. 

In this paper, we propose a control protocol for Scalable and Adaptive Information Dissemination (\name) in ICN,
a novel mechanism enabling large scale efficient data dissemination. We leverage the receiver-driven framework of NDN with
enhancements to overcome the out-of-sync problem.
\name achieves efficiency by separating reliability from congestion control.
By requesting for ``any'' packet that comes next, instead of the ``next packet in sequence'', \name is able to maximize the amount of packets delivered \emph{at the first attempt}.
Reliability is achieved by consumers performing information-centric repair which also has the attractive feature of preserving privacy and data integrity.
 While our design
is framed in the context of NDN, we believe \name can also be used by other reliable multicast solutions.
The contributions of this paper include:
\begin{itemize}
\item An analytical and emulation-based study on the out-of-sync problem that shows it will occur in real networks even with large in-network caches as long as there is receiver heterogeneity. This results in lower useful throughput, higher publisher and network load (see \S\ref{sec-out-of-sync}); 
\item A new reliable multicast framework that seeks to first maximize the useful throughput, by consumers requesting for \emph{Any Next Packet} (\delivery, see \S\ref{sec-basic-delivery}). Reliability is then achieved via a repair mechanism that leverages NDN's capability for receivers to request a data item from any network node including other receivers, while preserving privacy and data integrity (see \S\ref{sec-basic-repair});
\item A receiver-driven congestion control mechanism tailored for the \delivery that enables \emph{each} receiver to obtain its fair share of the bottleneck link while maintaining an application-efficient sending rate that doesn't necessarily slow down to the slowest
receiver's rate (see \S\ref{sec-congestioncontrol});    
\item Evaluations on our prototype show that \name is able to outperform ICP (by 50\%) and pgmcc (by almost 100\%) in terms of aggregate throughput without sacrificing network fairness; in a file content delivery application, \name can reduce aggregate network load (by up to $\sim$46\%) and transmission completion times (by more than 50\%) compared to ICP. Compared to pgmcc, \name reduces completion times by 40\% while only increasing network load by 10\% (see \S\ref{sec-eval}).


\end{itemize}

\section{Problem with Existing Congestion Control Mechanisms -- Out-of-Sync}
\label{sec-out-of-sync}
We first study
the out-of-sync problem in CCN and
show how it reduces the benefit of
in-network caches and the use of pending interests. We demonstrate this with an emulation using CCNx 0.8.0 along with a congestion control mechanism similar to ICP~\cite{ICP}.
We show through an analytical model that our observations on the
out-of-sync problem  is systemic with heterogeneous receivers, and
should be expected even with other receiver-driven, in-sequence congestion control mechanisms.
\hide{
\knote{leading the paper with this motivating foundation that of out-of-sync is the reason for our congestion control mechanism is fraught with risk and danger. Anyone who doesn't believe that out of sync is a major problem with a large enough cache will just dismiss the rest of the paper. Are we devoting too much for the out of sync problem and proving that it occurs that it ends up being the basis for the judgement of the paper rather than pointing out the value of a congestion and flow control mechanism that can support heterogeneous receivers?}
\jnote{I'm afraid if we only state we can deal with the heterogeneity of the receivers, the reviewers would say: existing congestion control mechanisms do not have to deal with it since there are caches in the network (all that they need to do is to add cache). We might have to tell them that cache in a deployable CCN cannot work well.}}

\subsection{Demonstration of out-of-sync in an emulated scenario}
\begin{figure}[t!]
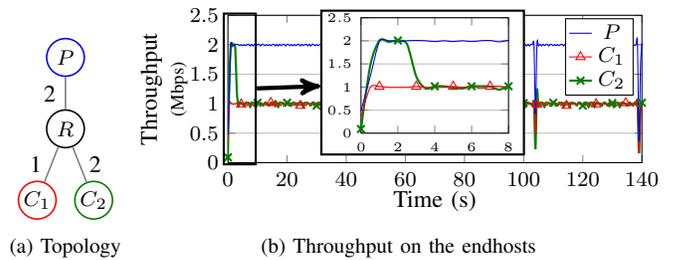

\centering
	\begin{minipage}[b]{0.2\linewidth}\centering%
    \input{SimpleTopology1.tex}
    \subcaption{Topology}%
    \label{fig:topology1}%
    \end{minipage}
	\begin{minipage}[b]{0.8\linewidth}\centering%
    \input{Result11.tex}
    \subcaption{Throughput on the endhosts}%
    \label{fig:result11}%
    \end{minipage}%
\caption{Out-of-sync problem emulated in a simple topology}
\label{fig:outofsyncsimple}
\end{figure}
\hide{
\begin{figure}[t!]
\centering
	\begin{minipage}[b]{0.695\linewidth}\centering%
    \input{EightSubTopology.tex}
    \subcaption{Dissemination tree}%
    \label{fig:topology2}%
    \end{minipage}\hspace{0.01\linewidth}%
	\begin{minipage}[b]{0.295\linewidth}\centering%
    \input{SallyFloydTopology.tex}
    \subcaption{Dumbbell}%
    \label{fig:topology3}%
    \end{minipage}%
\caption{Evaluation topology (bandwidth in $Mbps$)}%
\end{figure}}

To clearly demonstrate the out-of-sync problem and its cause, we use a simple emulation performed in Mini-CCNx. The network topology and the link rates (in $Mbps$) are shown in Fig.~\ref{fig:topology1}. The latency on all the links is $2ms$. Router $R$ has a 50 packets cache\footnote{We use a relatively small cache size to quickly demonstrate the out-of-sync problem.
However, this is fundamental and occurs even with much bigger caches.
Please see \S\ref{app-outofsync} for the relationship between the cache size and the heterogeneity allowed among the receivers.}.
\hide
{\knote{this has long been the problem - people are unconvinced with our statement that it is a meaningful problem with large caches, independent of how many times we say this.}
}
Consumers $C_1$ and $C_2$ start to request the same content
($\sim$$35MB$ in size, $8,965pkts$) from provider $P$ at the same time.
The throughput of the end hosts are shown in Fig.~\ref{fig:result11}.
For the first $2.5$ seconds, the PIT and the network cache benefit both receivers.
Requests from $C_1$ either get aggregated or get a cache hit at $R$. 
Overall network throughput in this period is $3Mbps$ (sum of downstream link capacities of $R$).
Subsequently, with heterogeneous receiver rates, the  receivers' requests
deviate farther apart.
The requests from $C_1$ can no longer be satisfied by the cache and
they are forwarded to $P$, to be treated as a distinct flow.
The response to these requests start to compete for the bandwidth on the link from $P$ to $R$ and the receive rate of $C_2$ is thus affected.
Since the congestion control protocol tries to achieve fairness between the receivers (flows),
the receive rate of the two consumers becomes $1Mbps$ each, and the overall network throughput reduces to $2Mbps$, thus under-utilizing the bandwidth by 33\%. 
For the entire transfer, we observed that $<2\%$ of the requests from $C_1$ see a cache hit. 

\begin{figure}[t!]
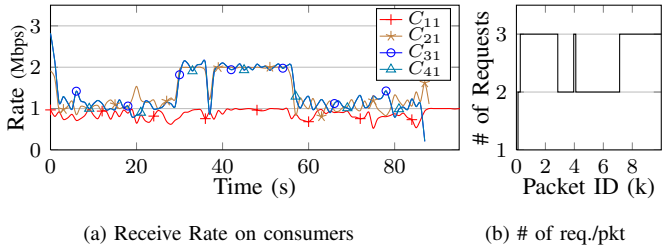

\centering
	\begin{minipage}[b]{0.66\linewidth}\centering%
    \input{Result32.tex}
    \subcaption{Receive Rate on consumers}%
    \label{fig:result22}%
    \end{minipage}\hspace{0.01\linewidth}
	\begin{minipage}[b]{0.33\linewidth}\centering%
    \input{Result31.tex}
    \subcaption{\# of req./pkt}%
    \label{fig:result21}%
    \end{minipage}%
\caption{Out-of-sync: On larger dissemination tree}
\label{fig:outofsynccomplex}
\end{figure}

Similar results occur in more complex topologies (\eg,
Fig.~\ref{fig:topology2} without dotted 
links).
The consumers $C_{11}$, $C_{21}$, $C_{31}$ and $C_{41}$ start to
request the same content from provider $P_1$ at the same time ($C_{51}$ is not active in this emulation).
The
cache size is $100pkts$ and the content size is $10,000$ packets.
The receive rates at the consumers are shown in Fig.~\ref{fig:result22} and the \# of requests (transmissions) per packet observed by $P_1$ is shown in Fig.~\ref{fig:result21}.
We can see that, similar to the pervious more simple case, the
consumers once
again get out-of-sync soon after the transmission sequence starts.
The \# of transmissions increase and the receive rate drops.
For the first 30 seconds, the aggregate throughput is only around
$4Mbps$ (the ideal throughput is $9Mbps$).
During the intervals $30$--$36$ and $40$--$55$ seconds, the faster
receivers get in-sync again, due to the randomness in the network and
cache occupancy.
The packet sequences between $2,800$--$3,950$ and $4,100$--$7,100$ are
transmitted twice and the throughput of $C_{21}$ through $C_{41}$
increases during these two time periods.
But even during this time, the aggregate throughput can only reach $7Mbps$ rather than the $9Mbps$ achievable in the ideal case.

Through these emulations, we see that due to the heterogeneity of the receivers, the cache in the intermediate routers might not be enough to absorb the difference in the request rates of the fastest and slowest receivers. This is the occurrence
of ``out-of-sync'' problem.
When the slower receivers re-issue requests, these requests are seen as being for a different ``new'' flow, since they can no longer be aggregated or be satisfied from the cache at the routers.
These ``new'' flows will then compete on the network links with packets of the original flow.
In some cases, this would even be with faster receivers on the common links, and affect their download rate as well.

The out-of-sync problem can happen even when all the receivers start their requests
for the sequence of data packets at the same time, even with the optimal cache replacement policy.
Note that the provider has to re-transmit packets as long as the intermediate router drops the packets within the gap (the difference in the sequence number of the packet requested by the fastest and slowest receiver).
\hide{\knote{huh? what is the gap?? the difference in the sequence number of the packet requested by the fastest and slowest receiver? Say it.}\jnote{added}.}
When the gap is larger than the available cache size for the flow, no matter which packet the replacement policy chooses, an additional transmission from the provider is required.


\subsection{Analytical model for out-of-sync occurrence}
\label{app-outofsync}
Receiver-driven feedback-based in-sequence congestion control protocols (\eg, TCP) share the following features:
\begin{inparaenum}[\itshape 1\upshape)]
\item each data consumer has a local view of the request as if he is the only consumer in the network,
\item all the data consumers tend to get a (statistically) fair-share of bandwidth, and
\item the packets in a data object are requested in-sequence and out-of-order is seen as an indication of congestion.
\end{inparaenum}
We realize that almost all the existing congestion control protocol proposed for CCN fall into this category.

To show the universality of the problem, we generalize the model for congestion control by assuming a best-case scenario where each receiver is receiving a flow of data with a constant bit rate which is exactly the fair-share that receiver can get.
We analyze the maximum heterogeneity that can be supported given a certain cache and flow size while the receivers still remain in-sync till the end of the flow.
Since we are focusing on a single flow, we also assume a simpler case that the network status (\ie, available bandwidth, cache size and latency) does not change during the lifetime of the flow.

We start with a more precise definition of out-of-sync.
\begin{DEF}[Out-of-sync]
\label{def-outofsync}
Consider a network with multiple routers interconnected in a tree topology and
a flow $f$ that has a provider on the top and receivers at the leaves of the tree.
At a branching router $N$, where available bandwidth to the downstream
receivers is in the range $[B_L,B_H]$, the out-of-sync occurs when the difference in the received file size (the gap, $G$) between the fastest and slowest receiver in the sub-tree below $N$ is larger than the available cache size C for flow $f$.
\end{DEF}

\begin{figure}
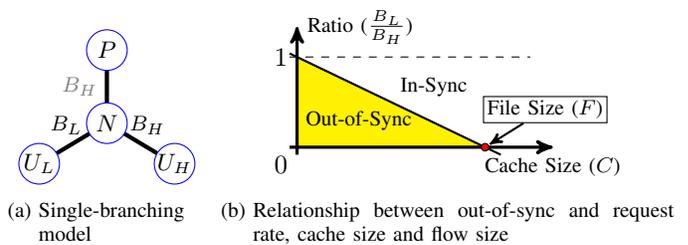

\centering
	\begin{minipage}[b]{0.3\linewidth}\centering%
    \input{SingleBranchTopology.tex}
    \subcaption{Single-branching model}%
    \label{fig-singlebranchingmodel}%
    \end{minipage}\hspace{0.02\linewidth}%
	\begin{minipage}[b]{0.68\linewidth}\centering%
    \input{SingleBranchData.tex}
    \subcaption{Relationship between out-of-sync and request rate, cache size and flow size}%
    \label{fig-singlebranchingdata}%
    \end{minipage}%
\caption{Out-of-sync in a single-branching model}
\label{fig-singlebranchingstructure}
\end{figure}

We then use a single-branch model to demonstrate the relationship among the cache size, flow size and receiver heterogeneity in a simple scenario.

\begin{LEMMA}
\label{lemma-outofsyncrequirement}
For a branching router $N$ with cache size $C$ in a dissemination tree, with the request rates of the immediate downstream links in range $[B_L, B_H]$, to avoid out-of-sync, the following condition should hold, \ie:
\begin{equation}
\label{equation-bwratioinsync}
\frac{B_L}{B_H}\geq(1-\frac{C}{F})\text{,}
\end{equation}
where 
$F$ is the size of the flow.
\end{LEMMA}
\begin{proof}
According to Def.~\ref{def-outofsync}, to avoid out-of-sync at $N$, it is sufficient to consider 2 immediate downstream links with the largest and smallest available bandwidth. That is, we can consider the single-branching topology in Fig.~\ref{fig-singlebranchingmodel} such that the
two data consumers ($U_L$ and $U_H$) are requesting for a same flow with size $F$ and their available bandwidth are $B_L$ and $B_H$ ($B_L$$\leq$$B_H$).

When the receivers are in-sync, the request sent upstream by $N$
targets a downstream rate of $B_H$, matching the receive rate of the
faster receiver (this is true for all protocols where
the network node does not perform an explicit congestion control
function and depends on the receivers to generate an appropriate request rate).
The download period for $U_H$ is:
$$t=\frac{F}{B_H}.$$
The maximum gap $G$ between the $U_L$ and $U_H$ is therefore:
$$G=(B_H-B_L)\times t=(1-\frac{B_L}{B_H})\times F$$
According to Def.~\ref{def-outofsync}, to keep the consumers in-sync, we need:
\begin{equation}
\label{equation-insync}
G=(1-\frac{B_L}{B_H})\times F \leq C.
\end{equation}
We can get (\ref{equation-bwratioinsync}) by reforming (\ref{equation-insync}).%
\end{proof}

The requirement for clients in-sync (equation \ref{equation-bwratioinsync}) is presented in Fig.~\ref{fig-singlebranchingdata}.
We can see that the requirement for being in-sync cannot be satisfied when the heterogeneity ($\frac{B_H}{B_L}$) is larger, the flow size $F$ is larger, and/or available cache size $C$ is smaller.

Since the request and data paths in large scale data dissemination usually form a tree structure rooted at the data provider, we then study the in-sync requirements in a $k$ level tree.

\begin{figure}
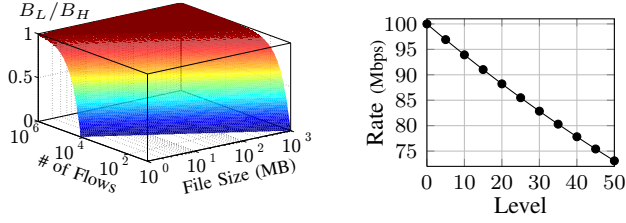

\centering
	\begin{minipage}[b]{0.53\linewidth}\centering%
    \input{3DPlot.tex}
    \subcaption{Relationship among out-of-sync criteria in real router}%
    \label{fig-3dplot}%
    \end{minipage}\hspace{0.01\linewidth}%
	\begin{minipage}[b]{0.46\linewidth}\centering%
    \input{RatevsLevel.tex}%
    \subcaption{Minimum allowed download rate \vs hierarchical level}%
    \label{fig-levelvsminrate}%
    \end{minipage}%
\caption{Difficulty of keeping heterogeneous receivers in-sync}
\label{fig-outofsync}
\end{figure}

\begin{THEOREM}
\label{theorem-outofsyncrequirement}
For a dissemination tree with $k$ levels and every intermediate router having cache size $C$,
all the receivers will be in-sync only when the available bandwidth between the fastest receiver and the slowest receiver follow:
\begin{equation}
\frac{B_L}{B_H}\geq(1-\frac{C}{F})^k
\end{equation}
\end{THEOREM}
\begin{proof}
We prove this theorem by contradiction.
Suppose that the request rates of the highest and the lowest receivers satisfy
\begin{equation}
\label{equation-insynchierarchicalwrong}
\frac{B_L}{B_H}<(1-\frac{C}{F})^k\text{,}
\end{equation}
and all the receivers are in-sync.

Without loss of generality, we assume that the receiver with the lowest rate $B_L$ is a downstream consumer of a router $N_t$ at level $t\in[1, k]$.
Let $B_{H,t}$ be the highest request rate among the downstream consumers of the router $N_t$.
Note that the consumer with request rate $B_H$ does not have to be the immediate next hop of $N_t$, the intermediate routers will always forward requests according to the fastest receiver.
According to Lemma~\ref{lemma-outofsyncrequirement}, we have
\begin{equation}
\label{equation-insynchierarchical2}
\frac{B_L}{B_{H,t}}\geq1-\frac{C}{F}\text{.}
\end{equation}
According to (\ref{equation-insynchierarchicalwrong}) and (\ref{equation-insynchierarchical2}), it follows that
\begin{equation}
\label{equation-insynchierarchical3}
B_{H,t}<B_H\times(1-\frac{C}{F})^{k-1}\text{.}
\end{equation}
The router $N_t$ is a downstream consumer of a router $N_{t-1}$ at level $t$$-$$1$.
Similarly, let $B_{L,t-1}$, $B_{H,t-1}$ be the lowest and highest rates among the downstream consumers of the router $N_{t-1}$ respectively. Since $B_{H,t}$$\geq$ $B_{H,t-1}$$\geq$$B_{L,t-1}$, according to Lemma~\ref{lemma-outofsyncrequirement}, we have
\begin{equation}
\label{equation-insynchierarchical4}
\frac{B_{H,t}}{B_{H,t-1}}\geq\frac{B_{L,t-1}}{B_{H,t-1}}\geq1-\frac{C}{F}\text{.}
\end{equation}
According to (\ref{equation-insynchierarchical3}) and (\ref{equation-insynchierarchical4}), it follows that
$$B_{H,t-1}<B_H\times(1-\frac{C}{F})^{k-2}\text{.}$$

By the similar argument, we can show that
\begin{equation*}
\begin{split}
B_{H,t-2}&<B_H\times(1-\frac{C}{F})^{k-3}\text{,}\\
B_{H,t-3}&<B_H\times(1-\frac{C}{F})^{k-4}\text{,}\\
~&~~\vdots~\\
B_{H,1}&<B_H\times(1-\frac{C}{F})^{k-t}\leq B_H\text{.}\\
\end{split}
\end{equation*}

Since the highest rate of downstream consumers of the router at level 1 should be $B_H$, \ie, $B_{H,1}$$=$$B_H$, we reach a contradiction and the proof is completed.
\end{proof}

From Theorem \ref{theorem-outofsyncrequirement} we can see that with the number of levels ($k$) increases, the gap between the fastest and the slowest receiver can become larger.
The design of the in-network cache helps in absorbing the heterogeneity of the receivers. 

Now we show that out-of-sync is difficult to avoid in a CCN router deployment.

\begin{remark}
The problem of receivers going out-of-sync persists with receiver-driven feedback-based in-sequence congestion control protocols as long as there
are  heterogeneous receivers.
\end{remark}

The cache size at a router will inevitably be much smaller than the total amount of content available in the network.
According to \cite{cachesize},  CCN requires a 25TB cache for a 50\% hit rate on Youtube data and 175TB cache for a 50\% hit rate on BitTorrent data.
But \cite{hardware} suggests that a deployable CCN router (with $\sim$\$1,500 overall hardware cost) would likely have around 100Gb of cache at current costs.
Thus, a router cache will be much smaller than the required cache size for
the kind of content accessed in current day networks.

For tractability, we assume that concurrent flows in a router share the 100 Gb cache size equally.
The relationship between the bandwidth ratio ($\frac{B_L}{B_H}$), file size ($F$) and number of flows is shown in Fig.~\ref{fig-3dplot}.
The intersection of the curve with XY plane is where $C$$=$$F$ (\# of flows$=$$\frac{100Gb}{F}$).
The region behind the curve represents the region receivers are out-of-sync.
Note that both the X- and Y-axis are log-scale, which means the area when
the receivers are in-sync is just a very small portion of the overall region.

Although a core router might have relatively larger cache, the number of concurrent flows on that router is also correspondingly large.
The available proportion of cache for each flow is therefore still quite small.
If a set of receivers request  20M bytes of data through a core router with 100k concurrent flows, the ratio of rates should satisfy the following:
$$B_L / B_H \geq 1 - (100Gb/100k) / 160Mb = 0.99375$$
If $B_H$ can reach 100Mbps, $B_L$ should be $\ge$99.3Mbps.
When we apply the hierarchical tree model in Theorem~\ref{theorem-outofsyncrequirement}, the minimum download rate \vs level is plotted in Fig.~\ref{fig-levelvsminrate}.
Even if we have 50 levels in such a hierarchy, the minimum required download rate is still $>$73Mbps.
This is difficult to achieve due to the number of
flows that may be multiplexed on a given link.

\hide{ 
\begin{IEEEproof}
Here, we prove the theorem in a recursive way by extending Lemma~\ref{lemma-outofsyncrequirement} into a $k$-level general tree topology.

\jnote{Do we still need it?}We first extend the topology in Fig.~\ref{fig-singlebranchingmodel} to a multi-receiver topology, \ie $N$ has downstream consumers $U_1$, $U_2$, \ldots, $U_n$. Since available bandwidth for the consumers are in the range $[B_L, B_H]$, Lemma~\ref{lemma-outofsyncrequirement} still holds. Therefore we only show the downstream nodes with highest and lowest rate as a representative of the siblings.

We then extend the consumer $U_L$ into an intermediate router ($N_L$) which behaves the same as $U_L$ from $N$'s point of view -- requesting with rate $B_L$ (shown in Fig.~\ref{fig-hierarchicalmodel}). According to the router behavior we described in Lemma~\ref{lemma-outofsyncrequirement}, the highest request rate under $N_L$ should be $B_L$. We denote the lowest request rate under $N_L$ as $B_{L2}$.

According to Lemma~\ref{lemma-outofsyncrequirement}, to keep in-sync on router $N_L$, the relationship between $B_L$ and $B_{L2}$ should follow:
\begin{equation}
\label{equation-insynclev2}
\frac{B_{L2}}{B_L}\geq(1-\frac{C}{F}).
\end{equation}
Combining (\ref{equation-bwratioinsync}) and (\ref{equation-insynclev2}), we can get:
$$\frac{B_{L2}}{B_H}\geq(1-\frac{C}{F})^2.$$

If we further extend $U_{L2}$ into $N_{L2}$ and perform the extension recursively, we can get the relationship between $B_{Lk}$ and $B_H$ as:
$$\frac{B_{Lk}}{B_H}\geq(1-\frac{C}{F})^k.$$

$B_{Lk}$ is the slowest receiver in the whole tree.%
\end{IEEEproof}
}
\hide{ 
\begin{figure}
\centering
	\begin{minipage}[b]{0.40\linewidth}\centering%
    \input{TreeTopology.tex}
    \subcaption{Generic hierarchical model}%
    \label{fig-hierarchicalmodel}%
    \end{minipage}\hspace{0.02\linewidth}%
	\begin{minipage}[b]{0.55\linewidth}\centering%
    \input{TreeExample.tex}
    \subcaption{Multi-level tree: numerical example}%
    \label{fig-hierarchicalexample}%
    \end{minipage}%
\caption{Hierarchical model to demonstrate Out-of-Sync problem}
\label{fig-hierarchicalstructure}
\end{figure}
}

\def\check{\cmark}
\def\uncheck{\xmark}
\begin{table}
\footnotesize
\centering
\caption{Overview of reliable multicast protocols}
\label{tab:protocols}
\begin{tabular}{@{\hspace{0.5mm}}c@{\hspace{0.5mm}}||@{\hspace{0.5mm}}c@{\hspace{0.5mm}}|@{\hspace{0.5mm}}c@{\hspace{0.5mm}}|@{\hspace{0.5mm}}c@{\hspace{0.5mm}}|@{\hspace{0.5mm}}c@{\hspace{0.5mm}}}
  & Network & Network & Application & App.-specific\\
  & Efficiency & Fairness & Efficiency & Reliability\\\hline
 SRM & \check  & \uncheck & \check & \check \\
 pgmcc & \check & \check & \uncheck & \uncheck \\
 CoolStreaming & \uncheck & \check & \check & \check\\
 ICP & \uncheck & \check & \uncheck & \check \\
\end{tabular}
\end{table}

\subsection{Existing solutions for preventing out-of-sync}
Existing NDN applications such as audio/video conferencing~\cite{VoCCN} seek to implement
in-sync delivery at the application layer. 
Initially, the communicating parties negotiate the data delivery rate (\eg, $n~pkt/s$) to maintain real-time communication, skipping packets arriving late.
Thus, instead of requesting the ``next-in-sequence'' packets, end hosts only request packets that can possibly arrive in time.

But this complicates the application, requiring it to negotiate the delivery rate a priori and also
continuously monitor packet arrival and consumption rates which may vary over time. The content
provider and consumer may find it difficult to set an appropriate sending rate given the network's
changing available bandwidth. Moreover the user experience may be sacrificed, with ``gaps'' or interruptions -
whether for conferencing or streaming content (\eg, video-on-demand(VoD)).
When the missing content
has to be retrieved (\eg, for VoD), this also results in out-of-sync behavior.


Thus, to maximize the utility of the network for file delivery and VoD, and to provide better support for real-time communications, there is a need for a transport-layer protocol that can keep the data receivers in-sync while retrieving data.

\section{Protocol requirements}
\label{sec-requirements}
To motivate our design, we sum up the requirements for an efficient
control protocol for large-scale data delivery and identify the issues
with some selected, existing solutions.
We categorize existing solutions and consider one representative (see Table~\ref{tab:protocols}): SRM~\cite{nak2} for reliable
IP multicast without congestion control; pgmcc~\cite{pgmcc} for
reliable IP multicast that aligns to the slowest receiver;
CoolStreaming~\cite{coolstreaming} for application-layer multicast,
and ICP~\cite{ICP} for a receiver-driven solution in ICN.

\begin{itemize}
\item[\textbf{Network Efficiency:}]
An efficient control protocol should maximize the utility of each
packet sent from the data provider, seeking to deliver
packets \emph{on the first attempt} to as many receivers as possible
  -- it is desirable keep the data consumers in-sync (where possible)
  when the data provider responds to their requests.  Receivers being
  out-of-sync causes higher overall transmissions, resulting in lower useful network throughput, higher
  aggregate network load and provider load.

\item[\textbf{Network Fairness:}]
Considering one-to-many flows, each flow should obtain a fair share
of the bottleneck bandwith 
on \emph{each path in the dissemination tree}.
  SRM injects data from the provider towards all the paths, ignoring
  this need to provide a fair share of the bottleneck capacity.

\item[\textbf{Application Efficiency:}]
We argue that a data provider sending at the slowest receiver's
consumption, like pgmcc is not application-efficient -- a very slow
receiver has an unduly significant influence on the application.
An alternative is to enable   
slower receivers to seek alternate means (\eg, get missing packets/repair  from other faster receivers).
 Repairs certainly add to the load on some segments of the network,
but the intent is to achieve a
 tradeoff between network load and session completion time.

\item[\textbf{Application-specific Reliability:}]
  Due to the difference in the requirements of the applications,
The protocol should be general and flexible to provide variable levels of
reliability based on application need. 
  \Eg, file delivery needs reliability without in-order delivery;
VoD would tolerate out-of-order delivery in a small range (within the
play out buffer size) and limited packet loss;
and conferencing applications do not need reliable delivery up to a point. 

\end{itemize}
As we show in Table~\ref{tab:protocols},  existing
approaches (characterized by the representative set) do not meet all these requirements, thus motivating our work on \name.

\section{\name Framework}
\label{sec-basic}


\begin{figure}[t!]
\centering
  \includegraphics[width=1\linewidth,natwidth=8.58,natheight=3.39]{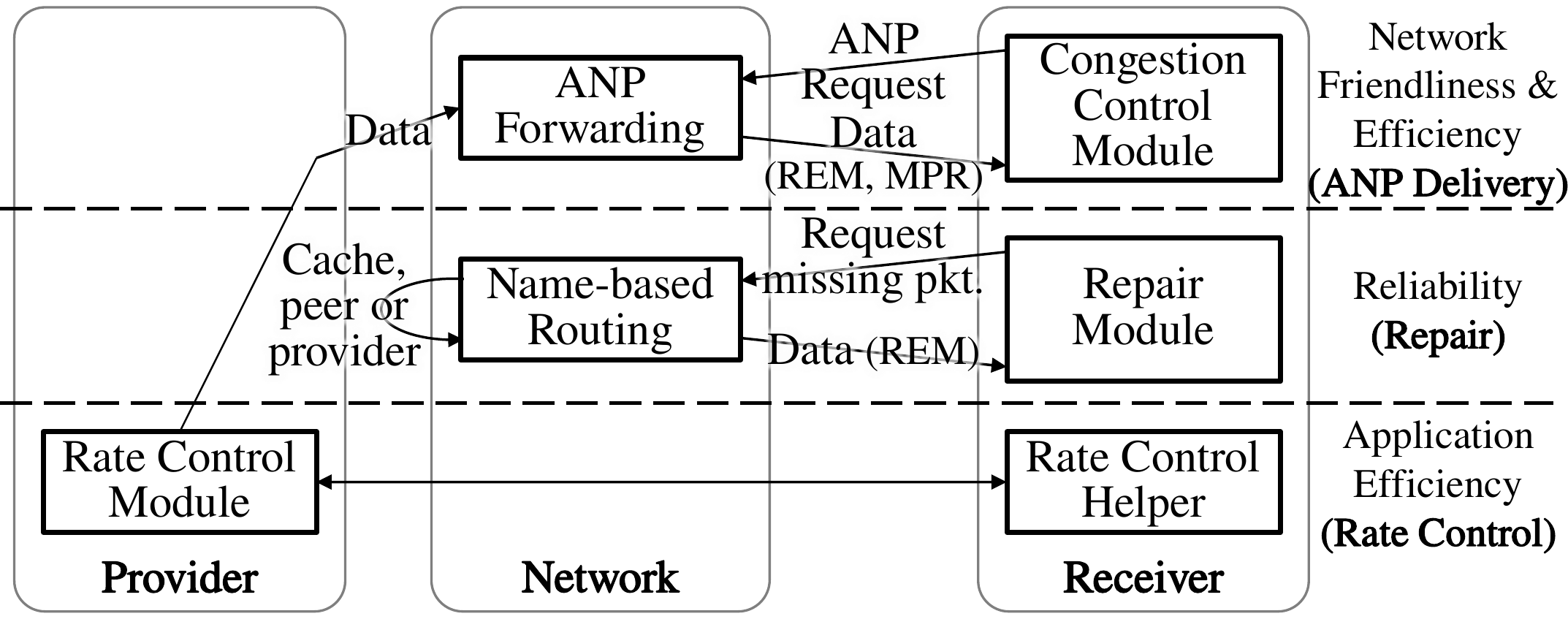}
  \caption{\name overview}%
  \label{fig:moduledivision}%
\end{figure}

Because the need for reliability and in-order delivery varies across applications, we argue that the one-many
(or many-many) information delivery and associated congestion control
mechanisms in multicast and ICN/NDN do not need to enforce in-sequence delivery and reliability by default.
Such solutions result in overheads (\eg, receivers being out-of-sync, with NDN/ICP) or inefficiency (sending
at the slowest receiver's rate, with pgmcc).
Therefore, we propose \name, a general framework for multiparty information delivery, that provides efficiency and
flexible reliability in the ICN/NDN context. The principles in \name can be generally applicable for multicast as
well. Fig.~\ref{fig:moduledivision} provides an overview of the proposed framework.


\name is designed to 
tolerate missing packets while maximizing the delivery probability of the packets on the first attempt (for \emph{network efficiency}), minimizing the need for retransmissions.
This is difficult to achieve in NDN's sequence-specific request model 
with varying network conditions and heterogeneous receiver capacities.
Therefore, receivers in \name use an \delivery model that requests for ``any-next'' packet and let the network
decide which packet can be delivered.

\emph{Reliability} can then be achieved by receivers leveraging NDN's sequence-specific request model to recover missing packets without sacrificing privacy and trust, since the request only needs to identify the
name of the missing packet and not necessarily the receiver's name or location (\name's repair).
Such repair requests can be satisfied by the content provider, caches or even other receivers.
Applications can prioritize the packets to be recovered, or even skip this. 

Many applications (\eg, file delivery) have elastic delay requirements.
The transmission rate of the provider chosen would be based on a tradeoff between the (network, provider) load and the content delivery completion time.
A higher sending rate reduces the completion time for the faster receivers, but could also result in higher retransmissions to reliably deliver to the slower receivers.
A lower sending rate may seek to operate at the opposite end of the spectrum -- sacrificing completion time for lower network load.
As a selective component at the application layer, \name accommodates a mechanism for the provider and consumers to negotiate the provider's sending rate.
\name allows the provider to pick a sending rate that ensures a application-dependent proportion
(\eg, 60\%) of the receivers can receive at that rate.
An \emph{ACKer} among the receivers is chosen to pace the flow, similar to pgmcc.
But, unlike pgmcc, the ACKer does not have to be the slowest receiver.
Although we do not cover the negotiation protocol for ACKer selection, due to space limitations (refer to our technical report~\cite{TechReport} for a more detailed description),  our evaluation of a file content delivery application in a real-world topology includes the ACKer selection protocol for controlling the sending rate (see \S\ref{sec-eval-filedelivery}).
\subsection{\delivery Model}
\label{sec-basic-delivery}

To overcome heterogeneous receivers getting out-of-sync, and maximize the likelihood of delivery of a packet in the first attempt, we propose a slight modification to the existing communication model of receivers asking for a specific packet (as is currently done by NDN and even TCP, \etc), whereby receivers ask for \emph{any} subsequent incoming packet.

We reuse the existing Interest packet in NDN for requests (\ie, prefix, selector and nonce in TLV format) with an extra field indicating this request is seeking ``any next Data'' that can satisfy the prefix in the Interest header.
In the CCNx based implementation we modify the \texttt{Exclude} field (to ``exclude'' all the previous Data available at the node).

On receiving \delivery requests, an upstream router does not perform the usual check of
the Content Store (cache) to avoid transmitting duplicate packets.
Instead, it places the prefix, the incoming face (as in NDN) and a pending request counter (PR) into its PIT.
PR is incremented on receiving an Interest packet with the same prefix and the routers propagate the maximum of the PR value upstream until this propagated value reaches the $1^{st}$ hop router of the data provider.
This modification is more space-efficient than the current sequence-specific requests.
For $n$ requests of the same flow, \name only needs 1 PIT entry and 1 counter compared to $n$ separate PIT entries with the existing approach.
When a router receives a Data packet of the flow (which satisfies the ANP request), it is forwarded
on the outgoing face and PR is decreased by 1.
The PIT entry will be removed when PR becomes 0.

With this subtle change, receivers in \name can get fresh data without getting out-of-sync, thus enhancing network efficiency.
This modification is also exploited for congestion control,  elaborated in \S\ref{sec-congestioncontrol}.


\subsection{Achieving Reliability via Efficient Repair}
\label{sec-basic-repair}
\hide{
With \name, we fill the required holes in the sequence via a repair mechanism.
ICN provides an efficient way to achieve sequence-specific data retrieval via the Pending Interest Table (PIT) and the Content Store.
ICN also provides a fundamentally important and desirable capability of ensuring \emph{privacy} of the
receivers, without the need to \emph{trust} other receivers (especially when issuing a request for repair from other receivers) compared  to the existing IP-based network.
In addition, the network has more information about the receivers downstream. We
optimize the Forwarding Information Base (FIB) propagation to redirect the repair requests towards nearby peer receivers who have already received those packets that are
missing at the requesting receiver.
This achieves lower network load and lower
content provider load as well.
These issues are covered in  \S\ref{sec-repair}.
}
For applications that need reliable delivery, we depend on the application interface to retrieve missing packets.
Applications should be able to determine which packets are of higher priority.
\name leverages NDN's sequence-specific data retrieval capability to retrieve the missing packets from any network node including peers. 



\subsubsection{Repair among Consumers}
\begin{figure*}[t!]
\centering
\begin{minipage}[b]{0.3\linewidth}\centering%
\input{LocalFIBPropagation2.tex}
\caption{Example of different repairs}%
\label{fig:localrepair}%
\end{minipage}\hspace{-0.01\linewidth}%
\begin{minipage}[b]{0.7\linewidth}\centering%
\input{ResultStateMachine.tex}
\caption{Result of window control at the receiver using state machine}%
\label{fig:resultwindow}%
\end{minipage}%
\end{figure*}

Similar to traditional NDN, in addition to the data provider, all the end hosts in \name that have (part of) the data can also support repair requests.
But, the difficulty lies in identifying and routing the requests to the receivers that have these missing data packets.
Moreover, since the repair packet has to share the bandwidth on the dissemination tree, independent of whether the repair request is delivered from a cache in the network or the provider, the repair has the potential to aggravate the congestion at a bottleneck link in
the path from the provider to the receiver.
Therefore, we propose a local FIB propagation optimization that can potentially mitigate the impact of repair on an already congested link.

With \name, after receiving (part of) the data from the provider, receivers will flood the prefix of the data over a limited number of hops. \Eg, in Fig.~\ref{fig:localrepair}, after receiving the packets, $C_1$ propagates prefix \texttt{/\conference/\name.pdf/ \_v1} over a 2 hop range.
When $C_2$ requests for \texttt{/\conference/ \name.pdf/\_v1/\_s20}, $R_7$ will forward the request to $R_6$ and eventually with get a response from $C_1$. The repair can then bypass the congested link $R_5$--$R_7$ and still increase the overall useful throughput to $C_2$.
We believe multiple paths will continue to exist, just as we see in current networks ~\cite{multipath1,multipath2,multipath3}.

Similar to peer repair solutions~\cite{nak2,rmx,frm,e2epubsub}, two major concerns with ``peer receiver-based repair''  are privacy for data consumers and the integrity of the data received from a peer\cite{nossdav}.
In IP (location-based) networks, the data consumer in most cases will reveal his own location (identity) while requesting from a peer (except those that explicitly choose to remain anonymous, \eg, by using Onion Routing).
NDN, however give these applications a natural way to achieve privacy and trust by default.
Since every data receiver requests repair via a Content Name, the receiver does not have to reveal his identity.
Even for peers that perform the local FIB propagation, they propagate the prefix of the flow or the chunk (a chunk represents $n$ packets) rather than their own identity.
Every Data packet in NDN has a key digest of the data provider and a signature for the data content from the provider.
The data consumer can easily check the integrity of the packet regardless of
whether it is received through repair from other receivers or from the original data provider.

\subsubsection{Prefix Granularity}
The simple mechanism we describe above may still be inefficient.
A receiver, say $C_1$ would propagate a FIB entry when it receives one or more packets of a flow so that repair requests for slower receivers can be redirected to $C_1$.
While propagating a FIB entry for each packet can result in excessive overhead,
propagating a single entry for the entire content
after receiving \emph{all} the packets of the flow may not be useful since $C_1$
can only respond to repair requests at the end of the flow.
To achieve a balance between the repair efficiency and reduced FIB size, we group
$n$ packets into a ``chunk'' and replace the \texttt{segmentID} in the ContentName with \texttt{chunkID/segmentID}.
\Eg, if $n$=100, the name of the packet with segmentID=205 would be \texttt{/\conference/\name.pdf/\_v1/\_c2/\_s205}.
Every packet still has a globally unique ContentName.
But $C_1$ can propagate \texttt{/\conference/\name.pdf/\_v1/\_c2} after receiving packets 200-299 for timely repair for other receivers.
The FIB entries created will be much smaller as well.

\section{Congestion Control for \delivery}
\label{sec-congestioncontrol}
\begin{figure}[t!]
    \centering
    \includegraphics[width=0.95\linewidth,natwidth=6.20,natheight=2.86]{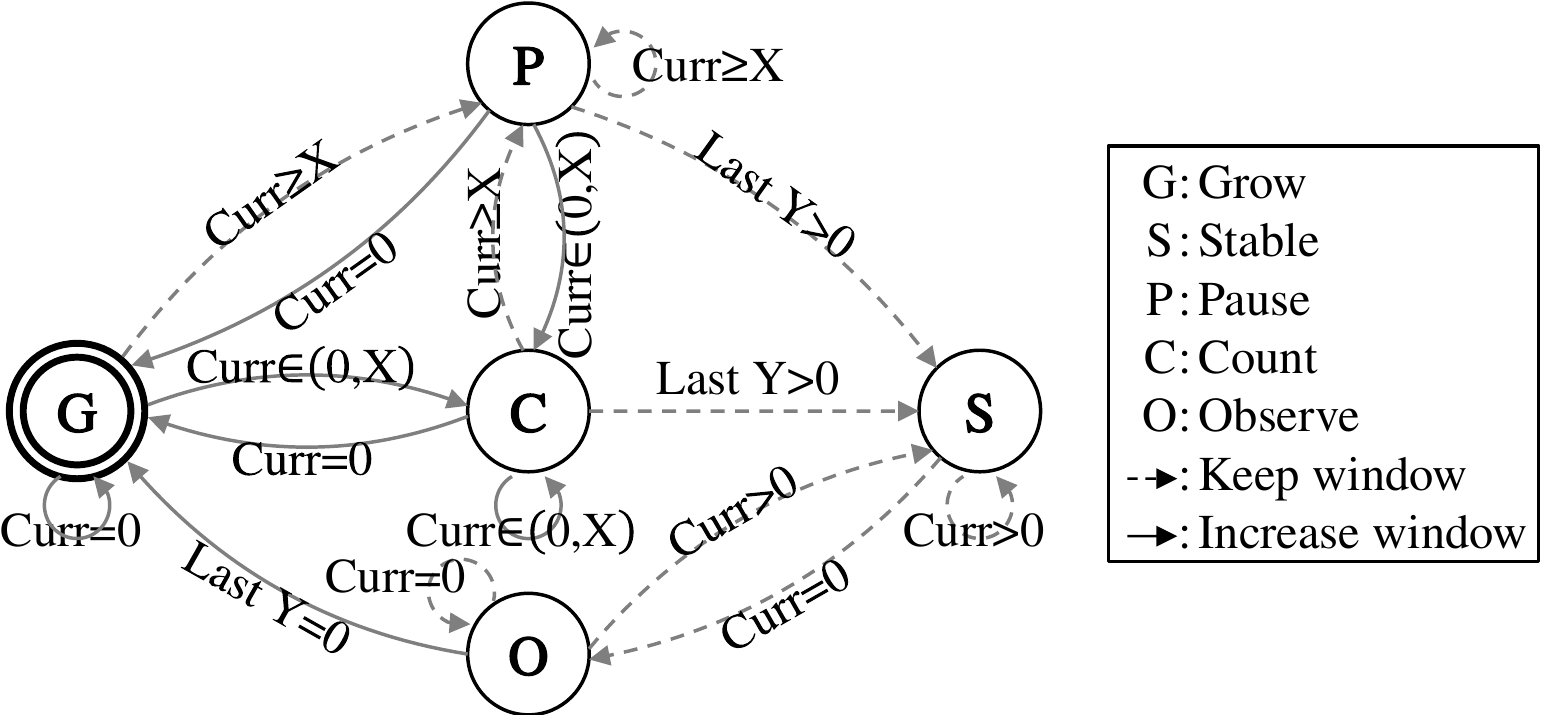}
    \caption{State machine for receiver window increase}
    \label{fig:window}
\end{figure}

A pure network based congestion control solution for multicast that divide sthe outgoing link bandwidth fairly for each flow going to the \emph{next hop}
is unable to accommodate indications of reduced demand from flows that have a limit downstream, especially with heterogeneous receiver bandwidths.
This results in inefficiency, as was suggested by~\cite{e2ecc}, and we show it in \S\ref{sec-eval-receiverlogic}
\name chooses to use an end-system assisted solution (receiver-driven) to avoid such inefficiency. 
First, \name retains the ``flow balance'' property suggested in NDN, in which every Interest packet will have at most one Data packet returned.
Receivers then maintain a window of unsatisfied ANP requests,
reflecting the maximum number of packets that can be in flight towards this receiver.
We can adapt this window, just as TCP, with an AIMD approach, to find the capacity from sender to receiver.
However, the sender's rate is not controlled by this window, when there are
heterogeneous receivers.
The difference between the sending and receiving rate naturally results in the receiver missing some of the packets in the sequence (we use the term ``holes in the sequence'').
These holes are not an indication of queue/buffer overflow (or congestion) on the routers,  and the receivers should not react accordingly.

We also make the important observation that because the sending rate is not controlled by the receivers' congestion windows, receivers with an available bandwidth higher than the transmission rate will not see congestion (or packet loss). Their window will continue to grow. 
This growth in the window essentially causes a large number of pending requests to be queued at the upstream router.
These pending requests will allow a large number of packets to be delivered in a burst towards the receiver (e.g., when the provider decides to increase the transmit rate)
This large build up of a pending requests queue at the bottleneck router
(with consequent increase in the feedback delay) causes that router to indicate
congestion (e.g., REM) for the large burst of packets.
This causes the receiver to react and reduce the window. This form of ``bang-bang'' effect has the potential to under-utilize the network and/or cause unfairness between flows for a subsequent interval.
While an excessively large window might cause inefficiency and unfairness, a very small window might also lead to inefficiency because a short burst caused by transient congestion might consume all the pending requests and result in unnecessary packet loss.
Because of these considerations, we develop additional mechanisms 
to achieve congestion control and fairness.

\subsection{Identifying Congestion \& Multiplicative Decrease}
\label{sec-congestioncontrol-REM}


We seek for help from the network to indicate congestion, using
Random Early Marking (REM~\cite{RED}). 
A simple AIMD mechanism (additive increase on receiving data and multiplicative decrease on seeing marks)
works acceptably on the receivers that have available bandwidth below the sending rate of the provider.
The introduction of REM into NDN can also simplify existing congestion control solutions and help avoid any unnecessary bufferbloat~\cite{bufferbloat}.


\subsection{Handling Additive Increase}
\label{sec-congestioncontrol-receivers}

\hide{
The multiplicative decrease on seeing REM can work well in the presence of competing flows, but a simple additive increase mechanism is not sufficient when the path to the receiver has high bandwidth.
Unlike TCP or ICP, the transmission rate on the provider is not controlled by the receivers' congestion windows.
The receivers with available bandwidth (or a fair share on a multiplexed link)
higher than the transmission rate will not see any marks and their window will continue to grow to a highly inflated value (especially if there is no limit).
This growth in the window essentially causes a large number of pending requests to be queued at the upstream router.
These pending requests will allow a large number of packets to be delivered in a burst towards the receiver when the provider decides to increase the transmission rate or new contention occurs on the path.
The large build up of a queue at the bottleneck router also increases the feedback delay.
Thus, for the period when a burst of packets is transmitted from a buffer that has high occupancy, those packets will be marked,
thus causing the receiver to react aggressively and reduce the window. This form of ``bang-bang'' effect has the potential to under-utilize the network or cause unfairness between flows over a
subsequent interval.
}

The optimal receiver window size can vary depending on the receiver, and can vary over time.
We propose a dynamic mechanism that observes and reacts to the instantaneous minimum pending request count (MPR) on the path, with a goal of keeping the
window size close to the optimal value.

The network piggybacks the instantaneous MPR value with data packets forwarded to receivers.
Note that different receivers can see different MPR values even for the same data packet since MPR is set at the branching point (at a router having other outgoing faces that can receive faster than this particular receiver).
A receiver observes the MPRs over a window so as to:%
\begin{inparaenum}[\itshape1\upshape)]
\item get the smallest value (thus avoiding excessive pending requests in the path) in that window,
\item ensure the minimum MPR (MMPR) is larger than a small value (to avoid all pending requests being consumed by a future short burst), and
\item smooth out variations.
\end{inparaenum}
To simplify the calculation, our approach decides on the window increase based
only on MMPR. It stops increasing the window when MMPR is larger than a configured
value $X$ and only increases window when it can rule out the possibility of reacting to a transient (based on the state machine depicted in Fig.~\ref{fig:window}, see~\cite{TechReport} for the detailed algorithm).

We use an example to trace the MMPR and the window increase decisions to aid in understanding how the scheme works (the value of $X$ in the state machine
is set to 5 for this example).
The MPR, window size and the state transitions in Fig.~\ref{fig:resultwindow} are shown for a receiver with $2Mbps$ available bandwidth.
Every cross represents a decision made at the end of a window.
The MMPR grows in accordance with the window size after $1 sec$. After 4 windows (decisions), the MMPR reaches 5 and the receiver stops increasing the window and MMPR stays around 5.
Although the MMPR becomes lower than 5 for several windows at around $4 sec$ (due to transient changes) the window size is not increased at the receiver.
When there is new contention at the bottleneck link (additional competing flow) occurring between $10 sec$ and $25 sec$, MMPR drops to 0 and the receiver then
begins to increase the window.
When the receiver receives marked packets (REM) at $12 sec$, it uses AIMD to respond to the
congestion introduced by the competing flow.
We show how this mechanism aids in achieving fairness in \S\ref{sec-eval-emulationfairness} and \S\ref{sec-eval-receiverlogic}.

\section{Evaluation}
\label{sec-eval}

\begin{figure*}[t!]
	\begin{minipage}[t]{0.055\linewidth}\centering%
    \begin{tikzpicture}
        \node[rotate=90,align=center] at (0,0) {\large{\bf \name}};
        \node[rotate=90,align=center] at (1em,0) {\small{Throughput}};
        \node[rotate=90,align=center] at (2em,0) {\scriptsize{(Mbps)}};
        \node at (0,-1.22) {};
    \end{tikzpicture}
    \end{minipage}
	\begin{minipage}[b]{0.27\linewidth}\centering%
    \input{CompetitionSAIDThroughput.tex}
    \end{minipage}
	\begin{minipage}[b]{0.2\linewidth}\centering%
    \input{CompetitionSAIDThroughput3.tex}
    \end{minipage}\hspace{1mm}%
	\begin{minipage}[b]{0.2\linewidth}\centering%
    \input{CompetitionSAIDThroughput4.tex}
    \end{minipage}\hspace{-3mm}%
	\begin{minipage}[b]{0.3\linewidth}\centering%
    \input{CompetitionSAIDThroughputA.tex}
    \end{minipage}%
    \\%
	\begin{minipage}[t]{0.055\linewidth}\centering%
    \begin{tikzpicture}
        \node[rotate=90,align=center] at (0,0) {\large{\bf ICP}};
        \node[rotate=90,align=center] at (1em,0) {\small{Throughput}};
        \node[rotate=90,align=center] at (2em,0) {\scriptsize{(Mbps)}};
        \node at (0,-1.76) {};
    \end{tikzpicture}
    \end{minipage}
	\begin{minipage}[b]{0.27\linewidth}\centering%
    \input{CompetitionICPThroughput.tex}
    \subcaption{$P_1$'s consumers}%
    \label{fig:competitionAA}%
    \end{minipage}
	\begin{minipage}[b]{0.2\linewidth}\centering%
    \input{CompetitionICPThroughput3.tex}
    \subcaption{$C_{31}$ \vs $C_{32}$}%
    \label{fig:competitionAB}%
    \end{minipage}\hspace{1mm}%
	\begin{minipage}[b]{0.2\linewidth}\centering%
    \input{CompetitionICPThroughput4.tex}
    \subcaption{$C_{41}$ \vs $C_{42}$}%
    \label{fig:competitionAC}%
    \end{minipage}\hspace{-3mm}%
	\begin{minipage}[b]{0.3\linewidth}\centering%
    \input{CompetitionICPThroughputA.tex}
    \subcaption{Aggregate}%
    \label{fig:competitionAD}%
    \end{minipage}%
\caption{Emulation result in competition scenario (using topology in Fig.~\ref{fig:topology2}, $X$=$10Mbps$)}
\label{fig:competitionA}
\end{figure*}

\hide{
In this section, we study the behavior of different components in \name with a custom simulator (which has been used in previous works~\cite{COPSS,CoExist}) and evaluate \name for both file delivery and streaming applications.
Two related mechanisms are compared with our solution:
\begin{inparaenum}[\itshape 1\upshape)]
\item A modified ICP solution that utilizes REM. While most of the proposed approaches use packet loss as an indication of congestion, we resorted to enhancing ICP with REM so as to have a reasonable comparison with \name in a network with heterogeneous receivers. 
\item pgmcc\cite{pgmcc} as an example of the solution that aligns the data provider transmission rate to the slowest receiver. It also uses a window-based congestion control to ensure the whole dissemination tree is fair to the bottleneck on the slowest path.
\end{inparaenum}
}

We first present evaluations based on our prototype implementation. We then present
simulation results for two applications (video streaming and large data file delivery)
on both a small custom topology (shown in Fig.~\ref{fig:topology2}) and a large RocketFuel Topology~\cite{Rocketfuel} with heterogeneous receiver bandwiths in our custom simulation environment (used in previous work~\cite{COPSS,CoExist}).

\textbf{Schemes compared:} 
\begin{inparaenum}[\itshape1\upshape)]
\item An ICP solution enhanced to use REM (so as to provide a reasonable comparison with \name).
\item pgmcc\cite{pgmcc}, a window-based congestion control protocol which aligns the data provider transmission rate to the slowest receiver. 
\end{inparaenum}

\subsection{Emulation of \delivery}

\begin{figure}[t!]
\centering
  \input{FiveSubTopology.tex}
  \caption{Dissemination tree topology (bandwidth in $Mbps$)}%
  \label{fig:topology2}%
\end{figure}

We implemented the core components: i) \delivery (described in \S\ref{sec-basic-delivery}); and ii) congestion control (described in \S\ref{sec-congestioncontrol}) in Linux\footnote{The implementation consists of roughly 1,500 lines of code and we make it publicly available in Github at \url{https://github.com/SAIDProtocol/SAIDImplementation.git}}. 
The modified forwarding engine is implemented in user space and packets are encapsulated in UDP, as with CCNx.

\textbf{Testbed and topology: }
Our testbed consists of \name enabled routers, each running on a single machine (Ubuntu 14.0, CCNx 0.4). All the end-hosts are run on a single machine to ease the collection of results (\eg, for clock synchronization reasons). The baseline topology (illustrated in Fig.~\ref{fig:topology2}) consists of 8 routers ($R_0$$\rightarrow$$R_7$), 7 consumers 
($C_{<x><y>}$)
and 2 providers ($P_1$, $P_2$).
Per link latency is 2ms and the numbers in the figure represent link bandwidth in Mbps. The bottleneck bandwidth for the consumers are marked in bold red.  

\subsubsection{Efficiency of \delivery Model}
\begin{figure}[t!]
\centering
    \input{4Sub.tex}
\caption{Aggregate throughput in 4-consumer scenario}
\label{fig:efficiency}
\end{figure}

\hide{
1.\jnote{Fig.~\ref{fig:efficiency}} same as 4-sub described in \S\ref{sec-out-of-sync}, $C_{11}$ 1Mbps, $C_{21}$ 2Mbps, $C_{31}$ 3Mbps, $C_{41}$ 3Mbps (sender is sending at 3Mbps), overall throughput see Fig.~\ref{fig:efficiency}, pgmcc overall: 1Mbps*4=4Mbps, conclusion: the \delivery allows for higher throughput during the first stage. (repair will be combined in application examples)
}

Fig.~\ref{fig:efficiency} illustrates the aggregate throughput achieved by \name and ICP in the presence of 4 consumers. pgmcc aligns to the slowest subscriber, and its results are easily obtained without
needing simulations\footnote{We implemented pgmcc in the simulation environment, but not the prototype environment}.  The experimental setup 
was used in \S\ref{app-outofsync} with consumers $C_{11}$,
$C_{21}$, $C_{31}$ and $C_{41}$, having bottleneck bandwidths of 1, 2, 3 and 4 Mbps respectively and the provider sending at 3Mbps. 
\name is able to achieve an aggregate rate close to the maximum achievable throughput (1Mbps ($C_{11}$) + 2Mbps ($C_{21}$) + 3Mbps ($C_{31}$) + 3Mbps ($C_{41}$)) of 9Mbps, while ICP's is around 4Mbps.
With ICP, receivers go out-of-sync and compete with one another on the link between $P_{1}$ and $R_{0}$. pgmcc can achieve approx. 4Mbps (1Mbps*4) throughput.

\hide{
2.\jnote{Fig.~\ref{fig:competitionA}} we have 5 consumers of p1 and 2 consumers of p2. c31, c32 are competing r2->r5, c41, c42 are competing r2->r6. X>1.5Mbps (we take 10 in eval) result of SAID and ICP shown in Fig.~\ref{fig:competitionA}
}

\subsubsection{Fairness in the presence of competition}
\label{sec-eval-emulationfairness}
The experimental setup is $P_{1}$ with 5 consumers (named $C_{<x>1}$ 
in
Fig. ~\ref{fig:efficiency} )
and $P_{2}$ has 2 consumers ($C_{32}$, $C_{42}$). To evaluate \name's fairness when there is a
competing flow from another provider ($P_{2}$), we configure $P_{2}$ to start sending packets after approximately 10s, with a sending rate of 3Mbps (\ie, as fast as it can) and study its influence in the network under an extreme case. $C_{31}$ and $C_{32}$ are competing for the bottleneck bandwidth between $R_{2}$ and $R_{5}$ (3Mbps) and $C_{41}$ and $C_{42}$ are competing for the bottleneck bandwidth between $R_{2}$ and $R_{6}$ (4Mbps). The bandwidth between $R_{5}$ and $C_{32}$, \ie X, is set to 10Mbps so that the link between $R_{2}$ and $R_{5}$ is the bottleneck link.

Fig.~\ref{fig:competitionA} illustrates the throughput achieved at each receiver. As in Fig.~\ref{fig:efficiency}, we observe in Fig.~\ref{fig:competitionAA} that in \name, each of $P_{1}$'s receiver is able to receive at the bottleneck capacity until 10s.
Once the competing flow arrives, the receivers in \name are able to receive at their \emph{(statistical) fair share} of the bottleneck bandwidth. In ICP, upon the arrival of a competing flow, the throughput achieved by each receiver drops significantly, from an average of 4Mbps to just 1Mbps.

Fig.~\ref{fig:competitionAB} and Fig.~\ref{fig:competitionAC} show that in \name, both $C_{31}$ and $C_{41}$ are able to receive at their individual fair share (1.5Mbps and 2Mbps respectively) in the presence of competition from $P_{2}$'s flow. On the other hand, with ICP, the consumer's of both $P_{1}$ and $P_{2}$ are not able to fully utilize their fair share of the bottleneck bandwidth, resulting in an under-utilization of the bottleneck bandwidths. This is due to the fact that both $C_{31}$ and $C_{41}$ are in fact competing with other consumer's from $P_{1}$ in the $P_{1}$ to $R_{0}$ link and $C_{32}$ and $C_{42}$ are competing with each other in the $P_{2}$ to $R_{0}$ link, due to the out-of-sync phenomena. The peak at the start of the $P_{2}$ flows in ICP, illustrate that for a short period, flows to $C_{32}$ and $C_{42}$ are in sync.

Finally, in Fig.~\ref{fig:competitionAD}, the aggregate throughput achieved by $P_{1}$'s and $P_{2}$'s consumers is close to the maximum achievable throughput with \name. The total throughput of the network in \name, up to 10s is close to the ideal 12Mbps. After 10s is even closer to the ideal 13Mbps.
With ICP however, it is approximately 5Mbps before 10s and 8Mbps afterwards. In pgmcc, in the absence of a competitor (\ie, before 10s), $P_{1}$ aligns to $C_{11}$ (1mbps) and therefore the total throughput is 5Mbps (since all its 5 consumers receive at 1Mbps each). In the presence of a competitor $P_{2}$ (after 10s) who is sending at 2Mbps (aligned to $C_{32}$), the total throughput will increase to 9Mbps.

\begin{figure}[t!]
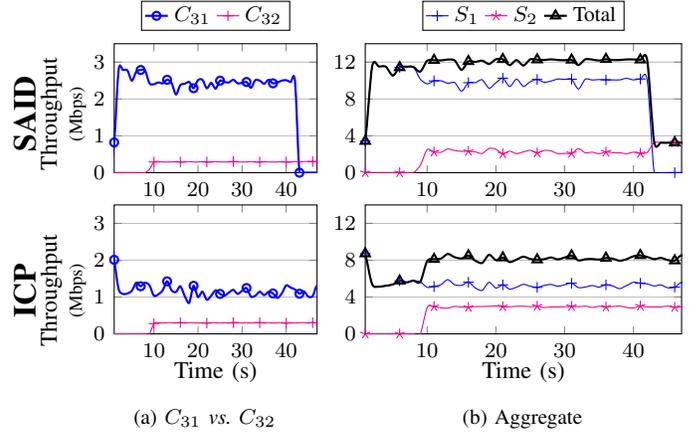

	\begin{minipage}[b]{0.12\linewidth}\centering%
    \begin{tikzpicture}
        \node[rotate=90,align=center] at (0,0) {\large{\bf \name}};
        \node[rotate=90,align=center] at (1em,0) {\small{Throughput}};
        \node[rotate=90,align=center] at (2em,0) {\scriptsize{(Mbps)}};
        \node at (0,-1.09) {};
    \end{tikzpicture}
    \end{minipage}%
	\begin{minipage}[b]{0.36\linewidth}%
    \input{CompetitionBSAIDThroughput3.tex}
    \end{minipage}%
	\begin{minipage}[b]{0.594\linewidth}%
    \input{CompetitionBSAIDThroughputA.tex}
    \end{minipage}%
    \\%
	\begin{minipage}[b]{0.12\linewidth}\centering%
    \begin{tikzpicture}
        \node[rotate=90,align=center] at (0,0) {\large{\bf ICP}};
        \node[rotate=90,align=center] at (1em,0) {\small{Throughput}};
        \node[rotate=90,align=center] at (2em,0) {\scriptsize{(Mbps)}};
        \node at (0,-1.75) {};
    \end{tikzpicture}
    \end{minipage}%
    \begin{minipage}[b]{0.36\linewidth}%
    \input{CompetitionBICPThroughput3.tex}
    \subcaption{$C_{31}$ \vs $C_{32}$}%
    \label{fig:competitionBA}%
    \end{minipage}%
	\begin{minipage}[b]{0.594\linewidth}
    \input{CompetitionBICPThroughputA.tex}
    \subcaption{Aggregate}%
    \label{fig:competitionBB}%
    \end{minipage}%
\caption{Emulation result in competition scenario (using topology in Fig.~\ref{fig:topology2}, $X$=$300kbps$)}
\label{fig:competitionB}
\end{figure}

\subsubsection{Benefit of receiver controlled mechanism}
We now modify our experimental setup 
to a simple dumbbell topology used in \cite{e2ecc} by changing the link bandwidth between $R_{5}$ and $C_{32}$, \ie, X to 0.3Mbps. Since $C_{32}$ can receive only at 0.3Mbps from $P_{2}$, $C_{31}$ is able to make use of the spare capacity of its bottleneck link between $R_{2}$ and $R_{5}$ and receive at approximately 2.7Mbps (See Fig.~\ref{fig:competitionBA}), thereby making optimal use of that link (total usage is 2.7Mbps + 0.3Mbps).
On the other hand, a flow based fair queuing approach would only achieve 1.8Mbps (1.5Mbps($C_{31}$)+ 0.3Mbps ($C_{32}$)), since the link between $R_{2}$ and $R_{5}$ would be shared equally between flows from $P_{1}$ and $P_{2}$, only for $P_{2}$'s packets to be dropped at $R_{5}$. This result highlights the benefit of using a receiver-driven approach for \name. In the case of ICP, $C_{31}$ is only able to receive close to an
equal
share of the $R_{2}$ and $R_{5}$ link bandwidth (<1.5Mbps). ICP gets approx 1.8Mbps
aggregate throughput on that link. The total throughout for \name (12 Mbps) and ICP (8 Mbps)
are shown in Fig.~\ref{fig:competitionBA}. For pgmcc, in the presence of a competitor, the total achievable throughput is 5.6Mbps (1Mbps * 5($C_{11}$-$C_{51}$) + 0.3Mbps *2($C_{32}$,$C_{42}$)).

\hide{
3.\jnote{Fig.~\ref{fig:competitionB}} same as 2, just X<1.5Mbps, we use 300kbps in the eval

\name:
c31 can use up the remaining bandwidth (3M-300k)bps, this is the benefit got from end-assisted congestion control; flow-based fair queueing can only achieve 1.5Mbps(c31)+300kbps(c32)=1.8Mbps compare to 3Mbps the ideal case (in our result). \name can still fully utilize the bottleneck bandwidth (ideal case 13Mbps in competition).

ICP:
link of R2->R5 under utilized (1Mbps(c31)+0.3Mbps(c32) < 3Mbps(BW)). But C42 is receiving faster since c32 is not that competitive. therefore, the total throughput can still reach around 8Mbps (1(c11)+1(c21)+1(c31)+1(c41)+1(c51)+0.3(c32)+3(c42))

pgmcc:
calculation, when competitor 1*5(c11-c51)+0.3*2(c32,c42)=5.6Mbps, really inefficient.
}

\label{sec-eval-receiverlogic}
\hide{
\begin{figure}[t!]
  \centering
  \includegraphics[width=0.3\textwidth]{figures/SallyFloydTopology}
  \caption{Demo topology for end-to-end congestion control}%
  \label{fig:topology3}
\end{figure}
}
\hide{
\begin{figure}[t!]
\centering
	\begin{minipage}[b]{0.495\linewidth}\centering%
    \input{MinorityFairness.tex}
    \subcaption{Receiver \vs Receiver}%
    \label{fig:followervsfollower}%
    \end{minipage}\hspace{0.01\linewidth}%
	\begin{minipage}[b]{0.495\linewidth}\centering%
    \input{MinorityFairnessWithQR.tex}
    \subcaption{Receiver \vs Query/Response}%
    \label{fig:followervsqr}%
    \end{minipage}%
\caption{Intra- \& Inter-protocol fairness of follower logic}
\label{fig:minorityfairness}
\end{figure}
}

\hide{
We illustrate the intra- and inter-protocol fairness of \name's receiver logic
(see Fig.~\ref{fig:minorityfairness}).
We use the topology in Fig.~\ref{fig:topology3}, with $X$ set as $10Mbps$. $C_i$ receives data from $P_i$ respectively.
$P_1$ sends at  $2Mbps$ and $P_2$ at  $3Mbps$.
Fig.~\ref{fig:followervsfollower}, illustrates \emph{intra-protocol fairness}.
All the receivers' receive rate is approximately $0.75Mbps$, thus achieving statistical fairness.
In Fig.~\ref{fig:followervsqr}, we illustrate the \emph{inter-protocol fairness} by changing the communication model between $P_1$ and $C_1$ to the original CCN/NDN communication. $P_2$ is sending at $2Mbps$ to $C_2$ using \name.
We can observe that during the co-existence period, both receivers are able to achieve statistical fairness (approx. $0.75Mbps$ each).

When $X$ is $128kbps$, similar result hold (but not shown). $C_2$ can receive at $128kbps$ and $C_1$ can receive the remaining.
This shows the benefit of an end-to-end congestion control. $C_1$ will only receive $0.75Mbps$ with hop-by-hop fair queueing.
}

\subsubsection{Multiple Bottlenecks: Fairness with \name}

\begin{figure}[t!]
  \centering
  \input{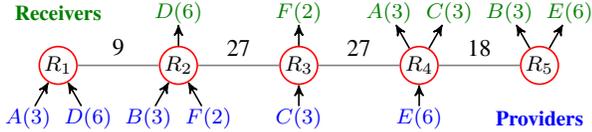}
  \caption{Generic Fairness Configuration (GFC)}\label{fig:topology4}
\end{figure}

We illustrate the fairness of \name's receiver-driven congestion control mechanism with a configuration
having multiple bottlenecks (Generic Fairness Configuration (GFC~\cite{GFC}))
so that different flows have different fair rates. The topology and the link capacities (in $Mbps$) is shown in Fig.~\ref{fig:topology4}. There are 6 flow groups ($A$$-$$F$) each with different \# of flows (in the bracket after the group name).
The ideal fair throughput of the flows 
are listed in Table~\ref{tab:fairness} (in the ``Fair'' column).

Since \name uses a receiver-side congestion window to control the receive rate, a feedback loop is formed between the receiver and a branching point. Thus the change in the congestion window of requests
does not affect the other receivers or the data provider.
Every ``flow'' in GFC can be seen as part of a dissemination tree (path from provider to receiver). The data providers
transmit data faster than any of the receivers' rate (at $27Mbps$).
After ignoring start up effects, we determine the average throughput of each of the flows in the subsequent  $15 sec$ period.

The average throughput and the fairness ratio ($\frac{avg.}{fair}$) for each flow with the two protocols are listed in Table~\ref{tab:fairness}.
Just like other feedback window-based congestion control mechanisms, \name slightly benefits the
flows with shorter RTT ($C$$-$$F$).
Similar results are observed for ICP.
In general, \name performs similarly to ICP (and all the other window-based congestion control mechanisms) from a fairness standpoint.
We believe \name would also be fair to  protocols using AIMD on \emph{each} path of the dissemination tree separately.

\begin{table}[t!]
\footnotesize
\centering
\caption{Fairness Result in GFC}
\label{tab:fairness}
\begin{tabular}{@{\hspace{0.5mm}}c@{\hspace{0.5mm}}|@{\hspace{0.5mm}}c@{\hspace{0.5mm}}|@{\hspace{0.5mm}}c@{\hspace{0.5mm}}|@{\hspace{0.5mm}}c@{\hspace{0.5mm}}|@{\hspace{0.5mm}}c@{\hspace{0.5mm}} @{\hspace{0.5mm}}c@{\hspace{0.5mm}}|@{\hspace{0.5mm}}c@{\hspace{0.5mm}} @{\hspace{0.5mm}}c@{\hspace{0.5mm}}}
\multirow{2}{*}{Group} & \multirow{2}{*}{Flows} & \multirow{2}{*}{Hops} & \multirow{2}{*}{Fair} & \multicolumn{2}{c|@{\hspace{0.5mm}}}{\name} & \multicolumn{2}{c}{ICP}\\\cline{5-8}
 & & & & Avg. & $\frac{\text{Avg.}}{\text{Fair}}$  & Avg. & $\frac{\text{Avg.}}{\text{Fair}}$ \\\hline\hline
 A & 3 & 5 & 1 & 0.7347 & 0.7347 & 0.7245 & 0.7245 \\\hline
 B & 3 & 5 & 2 & 1.2352 & 0.6176 & 1.2875 & 0.6437 \\\hline
 C & 3 & 3 & 6 & 6.5399 & 1.0900 & 6.4909 & 1.0818 \\\hline
 D & 6 & 3 & 1 & 1.1284 & 1.1284 & 1.1363 & 1.1363 \\\hline
 E & 6 & 3 & 2 & 2.3653 & 1.1826 & 2.3522 & 1.1761 \\\hline
 F & 2 & 3 & 9 & 9.0665 & 1.0047 & 9.0427 & 1.0047
\end{tabular}
\end{table}

\subsection{Simulation of different applications}
\hide{
\begin{figure}[t!]
\centering
	\begin{minipage}[b]{0.495\linewidth}\centering%
    \input{RepairEfficiency.tex}
    \subcaption{Receive rate on slow consumers ($P_1$ send at $6Mbps$)}%
    \label{fig:individual_repair_efficiency}%
    \end{minipage}\hspace{0.01\linewidth}%
	\begin{minipage}[b]{0.495\linewidth}\centering%
    \input{CompletionTimeVSSendingRateForRepair.tex}
    \subcaption{Average completion time with different sending rates}%
    \label{fig:repair_throughput}%
    \end{minipage}%
\caption{Repair efficiency}
\label{fig:repair_efficiency}
\end{figure}

To illustrate the efficiency of the repair mechanism, we simulated
a dissemination tree shown in Fig.~\ref{fig:topology2}.
$P_1$ is the provider. 
In addition to connectivity to $P_1$ over the binary dissemination tree, every data consumer also has a connection to a faster consumer with a
link between the corresponding first hop routers.
To limit the capability of such inter-router links, we set the bandwidth on these links to the bottleneck bandwidth that the receiver has.
\Eg, $C_1$'s bottleneck bandwidth is $1Mbps$, therefore the link between $R_7$ and $R_8$ is also $1Mbps$.

\subsubsection{Individual throughput with repair}

In this simulation, $P_1$ sends a total of $120Mb$ data to all the receivers at $6Mbps$ (a balance between
the receive rates of different receivers). 
The instantaneous throughput on $C_1$ to $C_3$ is shown in Fig.~\ref{fig:individual_repair_efficiency}.  $C_1$ has a $1Mbps$ bottleneck from $P_1$. For the first 1.5 seconds,  $C_1$'s throughput is $1Mbps$. After that, $C_1$ can get repairs from faster receivers, and its throughput reaches $2Mbps$. 
The throughput on $C_2$ shows almost the same trend.
But since $C_1$ and $C_2$ share the link between $R_8$ and $R_9$, the repair from $C_3$ to $C_1$ affects the performance of $C_2$.
The overall throughput on $C_2$ is only $3Mbps$ when it cannot satisfy all the requests from $C_1$ within the first 10 seconds.
The throughput of $C_4$ and $C_5$ are close to $6Mbps$ since they can get $4Mbps$ and $5Mbps$ from the provider directly and the remaining from $C_6$.

\subsubsection{Overall efficiency under different sending rates}

We then vary the sending rate on the provider and measure the overall completion time for three cases: repair via provider, peer repair and ICP.
The average completion time is shown in Fig.~\ref{fig:repair_throughput}.
We see that when the data provider aligns the sending rate to the slowest receiver ($1Mbps$, like pgmcc), no repair is needed. But the average completion time is even higher than ICP (as shown in Fig.~\ref{fig:result22}, most of the receivers can receive at around $2Mbps$). 
When the sending rate becomes higher, the average completion time reduces.
The peer repair solution helps receivers get repair packets even before the flow is finished. Therefore it always has a lower average completion time than the ``repair via the provider'' solution. The completion time reduces below ICP when the sending rate is above $2Mbps$,
with reduced provider load (\name retransmit rate is less than 2 compared to 3.36 for ICP).
As the sending rate increases, the peer repair reduces the load on
the provider, compared to both of the other solutions, albeit not necessarily
reducing the average completion time significantly.
}

\subsubsection{Video Streaming}
\begin{table}[t!]
\footnotesize
\centering
\caption{\Pt ($s$) in streaming demo (video length=$40s$)}
\label{tab:streaming}
\begin{tabular}{@{\hspace{1mm}}r@{\hspace{1mm}}|@{\hspace{1mm}}r@{\hspace{1mm}} @{\hspace{1mm}}r@{\hspace{1mm}} @{\hspace{1mm}}r@{\hspace{1mm}} @{\hspace{1mm}}r@{\hspace{1mm}}|@{\hspace{1mm}}r@{\hspace{1mm}}}
           & $C_1$ & $C_2$  & $C_3$ & $C_4$ & Rep. \\\hline\hline
  Baseline & 83.384 & 22.507 & 2.461 & 0.886 & --\%  \\
  ICP & 90.530 & 33.965 & 33.821 & 33.820 & --\% \\\hline
  pgmcc & 84.804 & 84.770 & 84.768 & 84.767 & 0.00\% \\\hline
  \name-F    & 83.821 & 40.062 & 39.569 &  1.131 & 12.44\%  \\
  \name-S  & 83.541 & 22.754 & 4.010 &  1.123 &  21.91\% \\
  \name & 44.304 & 1.271 & 1.151 & 1.131 & 12.44\% \\
\end{tabular}
\end{table}
\hide{
\begin{figure}[t!]
\centering
	\begin{minipage}[b]{0.45\linewidth}\centering%
    \input{StreamingResultQR.tex}
    \subcaption{Baseline(g) \vs NDN(b)}%
    \label{fig:streamingqr}%
    \end{minipage}\hspace{0.01\linewidth}%
	\begin{minipage}[b]{0.54\linewidth}\centering%
    \input{StreamingResultPS.tex}
    \subcaption{\name-F(gray) \vs \name-S(black)}%
    \label{fig:streamingps}%
    \end{minipage}%
\caption{Repair efficiency for streaming with different video size}
\label{fig:streaming}%
\end{figure}
}

We consider a streaming video application with a playout rate of
$3Mbps$.  We evaluate the effectiveness of the approaches using the
\emph{\Pt}, which reflects the impact on user experience.
No stalls
occur when there are no holes in the sequence in the play out
buffer for the next $1s$ of video playback.
For the baseline, we run the simulation $4$ times with a single
receiver requesting the video.
\name-F is when repair is by the provider at the end of the flow (similar to ``download-and-play'').
\name-S is repair by the provider while streaming the video.
\name is the peer-assisted repair solution we propose here.
These solutions are compared on the tree topology (Fig.~\ref{fig:topology2}, with only $C_{11}$-$C_{41}$ activated) as well as ICP and pgmcc.

Table~\ref{tab:streaming} shows the \pt and the repair ratio
($\frac{\text{\# of pkts via repair}}{\text{\# of total packets}}$, 'Rep' in Table) for a $40s$ video.
In ICP, when all receivers request the video simultaneously,
they go \emph{out-of-sync} soon thereafter. The \pt becomes larger than the baseline especially for faster receivers. 
With pgmcc, since the provider has to align with the slowest receiver
($C_1$), the \pt for the rest of the receivers is up to $80s$ larger compared to the
baseline. The user experience for the faster receivers therefore deteriorates considerably.

Although \name-F achieves a relatively low repair rate, the \pt for $C_1$-$C_3$ remains high, because the repair is performed after the flow finishes.
$C_2$ and $C_3$ benefit from the repair during streaming in \name-S.
But this benefit comes at the cost of higher network load.
The repair ratio of \name-S is higher than \name-F ($\sim$$22\%$ \vs $\sim$$12.5\%$) since the retransmission has to go through the bottleneck link and affects the primary ``any packet'' stream.
But the retransmission rate from the provider of \name-S is still lower than ICP ($\sim$$1.9$ \vs$\sim$$3.4$, not shown) which means \name-S consumes less network and provider resources.
\name (our proposal) however is superior as it is able to utilize the extra bandwidth between end-hosts.
The repair does not affect the multicast session and the slower consumers can get twice the bandwidth compared to \name-F and \name-S. Despite the repair, the \pt for the slower receivers ($C_1$$-$$C_3$) is much smaller than with the other solutions, even though they are playing the video at the same rate of $3Mbps$.
We varied the video length and observed that the \pt grows proportionally with the video length but the pattern in Table~\ref{tab:streaming} still holds.  


\hide{
\mnote{Move part of this to summary}
From the results, we see that separating the flow and congestion control from the in-sequence reliable delivery enables us to achieve a higher overall throughput (and lower average completion time), lower network load and lower provider load. In CCN, the reliability in data transmission is no longer the sole responsibility of the sender as in unicast, location-oriented protocols. The data consumers can also help each other in the reliable data delivery.
}
\hide{
\subsection{Evaluation of ACKer Selection}

\begin{figure}[t!]
\centering
    \begin{minipage}[b]{0.255\linewidth}\centering%
        \input{ACKerTopology2.tex}
        \subcaption{Topology}%
        \label{fig:ACKerTopology}%
    \end{minipage}
  	\begin{minipage}[b]{0.745\linewidth}\centering%
    \input{ACKerCompetitor2.tex}
        \subcaption{Rate of $P_1$'s receivers in competition demo}%
        \label{fig:ackershiftcompetitor}%
    \end{minipage}\\
  	\begin{minipage}[b]{0.495\linewidth}\centering%
        \input{ACKerReceiversJoin.tex}
        \subcaption{Rate \& ACKer on receivers join}%
        \label{fig:ackerreceiverjoin}%
    \end{minipage}\hspace{0.01\linewidth}%
  	\begin{minipage}[b]{0.495\linewidth}\centering%
        \input{ACKerReceiversLeave.tex}
        \subcaption{Rate \& ACKer on receivers leave}%
        \label{fig:ackerreceiverleave}%
    \end{minipage}%
  \caption{Evaluation results for ACKer selection}
\end{figure}

\hide{
\begin{figure}[t!]
\centering
    \begin{minipage}[b]{1\linewidth}\centering%
        \input{ACKerTopology2.tex}
        \subcaption{Evaluation topology}%
        \label{fig:ACKerTopology}%
    \end{minipage}\\%
  	\begin{minipage}[b]{0.495\linewidth}\centering%
        \includegraphics[width=1\textwidth]{figures/ACKerCompetitor}
        \subcaption{Rate of $P_1$'s receivers in competition demo}%
        \label{fig:ackershiftcompetitor}%
    \end{minipage}\hspace{0.01\linewidth}%
  	\begin{minipage}[b]{0.495\linewidth}\centering%
        \includegraphics[width=1\textwidth]{figures/ACKerTwoFlow}
        \subcaption{Rate of $P_1$ \& $P_2$ in two-flow demo}%
        \label{fig:ackertwoflow}%
    \end{minipage}\\%
  	\begin{minipage}[b]{0.495\linewidth}\centering%
        \includegraphics[width=1\textwidth]{figures/ACKerReceiversJoin}
        \subcaption{Rate \& ACKer on receivers join}%
        \label{fig:ackerreceiverjoin}%
    \end{minipage}\hspace{0.01\linewidth}%
  	\begin{minipage}[b]{0.495\linewidth}\centering%
        \includegraphics[width=1\textwidth]{figures/ACKerReceiversLeave}
        \subcaption{Rate \& ACKer on receivers leave}%
        \label{fig:ackerreceiverleave}%
    \end{minipage}%
  \caption{Evaluation results for ACKer selection}
\end{figure}

}

We use a synthetic topology (Fig.~\ref{fig:ACKerTopology}) to demonstrate the correctness of the ACKer selection logic. 10 consumers $S_1$-$S_{10}$
are on the $1^{st}$ hop router with two data providers $P_1$ and $P_2$
on the other side. Consumer $S_i$ has an available bandwidth of $iMbps$. Both the publishers have enough bandwidth to support the fastest receiver.
For the data providers, we set the transmit rate requirement to
be in the bandwidth range of $55\% \pm 5\%$. This
places a corresponding requirement on the selection of the ACKer
(without competition $S_6$ should be the ACKer). The receive statistics are collected once every 2 seconds.

\subsubsection{Competition on the ACKer}

Receivers $S_1$ to $S_{10}$ receive data from provider $P_1$, and $S_6$ receives data from both providers $P_1$ and $P_2$. We manually assign a receiver with $6Mbps$ bandwidth 
as $P_2$'s ACKer.  
The receive rates observed on $S_4$, $S_5$, $S_6$ and $S_{10}$ are shown in Fig.~\ref{fig:ackershiftcompetitor}. $P_1$ starts first.
We see that in the first round, the provider picks $S_{10}$ as the ACKer since his reply reaches the provider first.
From the second round onwards, the provider picks the correct ACKer based on our settings -- the maximum receive rate across the consumers is $6Mbps$.
Provider $P_2$ starts sending at $8s$. 
$P_1$ rate drops to $3Mbps$ (fair share) at that point. 
the receivers' rate.
At the next round of ACKer selection (starting at $10s$),
$P_1$ realizing $S_6$ is no longer eligible. Based on the statistics, he picks $S_5$ as the new ACKer.
When $P_2$ finishes at $14s$, the receive rate of $S_6$ grows back to the sending rate ($5Mbps$). Since $S_6$ sees fewer marks, the ACKer selection
done at $18s$ causes $P_1$ to shift back to $S_6$.

\hide{
\subsubsection{Two flow competition}
\mnote{Remove this subsection and the corresponding figures to obtain space}\jnote{agree. and we can save space for a figure. But need to mention in the prev subsubsection, when the two-flow is fair, they can also achieve fair results. we omit it due to the lack of space.}
The second example we show here is the fairness when two flows have the same network condition. We reuse the topology shown in Fig.~\ref{fig:ACKerTopology}, but all the receivers receive a flow from $P_1$ and $P_2$ each. $P_1$ starts sending at time $0s$ while $P_2$ starts sending at around $8s$. The sending rate of the providers are shown in Fig.~\ref{fig:ackertwoflow}.

The ACKer shift events are not shown in the figure due to the lack of space. We observed that, $P_1$ shifts to $S_7$ at $8s$ when $P_2$ started and picked $S_{10}$ as ACKer. From the second round on, $P_2$ chooses $S_6$ as its ACKer. In the following round of $P_1$, $S_6$ is again elected as ACKer. The sending rate shows the statistical fairness between the two providers. The fluctuation appears due to the probability of marking from REM. During the period when the two flows share bandwidth, the average sending rate is around $5Mbps$ from $P_1$ and $3Mbps$ from $P_2$.
}

\subsubsection{Tolerating Receiver Joins and Leaves}


Receivers joining and leaving (\ie, churn) might also affect the
selection of a new ACKer.
$S_1$ is the only receiver at the beginning. After $3s$, a faster receiver joins every 6 seconds ($S_i$ joins at $(i$$-$$2)$$*$$6$$+$$3s$). For leave events, $S_1$ to $S_{10}$ are receiving the flow. From $3s$ onwards, a fast receiver leaves every 6 seconds ($S_i$ leaves at $(10$$-$$i)$$*$$6$$+$$3s$). The sending rate and the selected ACKer are shown in Fig.~\ref{fig:ackerreceiverjoin}-\ref{fig:ackerreceiverleave}. While the receivers change, the data provider can pick a proper ACKer as needed. As long as the original ACKer is within the required
bandwidth range, the provider does not pick a new ACKer.
}

\subsubsection{File Content Delivery}
\label{sec-eval-filedelivery}
\begin{figure*}[t!]
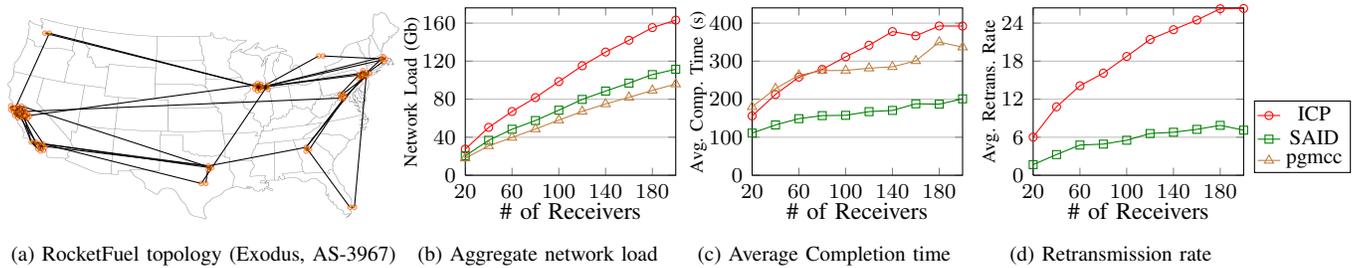

  	\begin{minipage}[b]{0.30\linewidth}\centering%
\input{rocketfuel.tex}
         \subcaption{RocketFuel topology (Exodus, AS-3967)}%
       \label{fig:rocketfuel}%
    \end{minipage}\hspace{-0.01\linewidth}%
  	\begin{minipage}[b]{0.21\linewidth}\centering%
        \input{OverAll_Network.tex}
        \subcaption{Aggregate network load}%
        \label{fig:overall-network}%
    \end{minipage}%
  	\begin{minipage}[b]{0.21\linewidth}\centering%
        \input{OverAll_CompletionTime.tex}
        \subcaption{Average Completion time}%
        \label{fig:overall-completion}%
    \end{minipage}%
  	\begin{minipage}[b]{0.21\linewidth}\centering%
        \input{OverAll_Retransmission.tex}
        \subcaption{Retransmission rate}%
        \label{fig:overall-retransmission}%
    \end{minipage}%
    \begin{minipage}[b]{0.05\linewidth}\centering
    \ref{legend:overall} \\~\\~\\~\\
    \end{minipage}%
    \caption{Simulation result for file content delivery application}
    \label{fig:overall}
\end{figure*}

We evaluated the complete \name solution with the RocketFuel topology (Fig.~\ref{fig:rocketfuel}). We randomly place 20-200 receivers on the 79 core routers in the topology.
Since RocketFuel does not have bandwidth information, we assign available bandwidths
in the range of $1$-$10Mbps$ for each link.
The result of a trace with 100 flows using ICP, \name and pgmcc is in Fig.~\ref{fig:overall}.

By decoupling  reliability from  congestion control, \name has lower network load compared to 
ICP (Fig.~\ref{fig:overall-network}), especially as the number of receivers increases,
by up to $46\%$ with 200 receivers.
Since $\sim$$60\%$ of the data is delivered at the first attempt, \name has a much lower average \# of transmissions of each packet from the provider and average flow completion time compared to ICP (Fig.~\ref{fig:overall-retransmission}).
But \name consumes more network bandwidth ($\sim$$10\%$) compared to pgmcc
since it aligns to a faster receiver and uses repairs
subsequently. For the $10\%$ additional network load, \name achieves lower average completion time (by $\sim$$65\%$, Fig.~\ref{fig:overall-completion}).
%

\section{Related Work}
\label{sec-motiv}
Here we focus on the related work that are not mentioned earlier. 

Many reliable multicast protocols have been proposed to enable large scale reliable data dissemination.
To maximize the utility of multicast, cyclic- and scheduled-multicasts~\cite{cyclicmulticast,scheduledmulticast} have been proposed to benefit the consumers that are not starting at the same time.
These solutions can be broadly classified into two categories based on the chosen repair mechanism:
provider repair or peer-assisted repair.

The clients in provider-repair mechanisms~\cite{gobackN,satellitebc,nak1,nak3,mcforhighspeed,norm} send NAKs to the provider (or the whole group to suppress duplicate NAKs) and the provider retransmits the missing packets specified in the NAKs.
The data provider in such solutions has to align the sending rate to the slowest receiver eventually.

Peer-assisted approaches like \cite{nak2,rmx,frm,e2epubsub,rmtp,tram} group receivers in a hierarchical structure and the receivers ACK to the upper level in the tree so that the ACKs can be aggregated.
By introducing such a relationship among receivers, these solutions allow receivers to perform local repair and therefore, the provider can align the sending rate to the majority of (or the fastest) receivers based on the ACK strategy.
Unfortunately, in these proposals, the subscribers have to exchange information in a peer-to-peer manner in the IP network to perform repair as well as send ACKs.
According to~\cite{nossdav}, these solutions face the problem of \textbf{privacy} and \textbf{trust}.
Namely, the receivers have to reveal their identities (IP addresses) to the peers, and thereby trust them as there is no guarantee of data integrity during peer-repair.

Layered multicasts are also proposed to deal with heterogeneous consumers.
In \cite{layered1,layered2}, the provider creates different multicast groups that transmit different resolutions of the data.
The receivers can select appropriate groups according to their link capacity.
These solutions are applicable to select applications.
But having a reliable multicast capability for a single rate stream is still fundamental for broad-based use across all kinds of applications.
Therefore, the layered multicast solutions can be considered as
orthogonal to the single-rate reliable multicast.


\hide{
NDN introduces a new forwarding engine model which is composed of an FIB, a PIT and a Content Store. 
The PIT can aggregate Interests 
that arrive within a short time period for the same piece of data from different requesters.
On receiving a Data packet, the forwarding engine at a router will send the packet to all the interfaces (generalized to faces in NDN) that have pending interests.
Content Store caches all the data packets that come through and stores them as long as possible to
satisfy future requests.
}
Many congestion-control mechanisms have been proposed for ICN like\cite{ICP,hrtcp,ictp,contug,cctcp}. They all face the out-of-sync issue since they rely on similar TCP-like mechanism.
Other works such as \cite{pursuit,COPSS} try to achieve efficient large scale data dissemination via pub/sub. 
But they lack an efficient mechanism to ensure reliability and avoid congestion collapse in the network. 

\hide{
\section{Discussion}
Very important: what happens when the receivers are not starting from the same time?

They can still use the mechanism and see the previous parts as the missing parts and do repair.
Or, solutions like cyclic-multicast, wait and publish can be adopted.

DASH

\jnote{Need to mention: $X$ and $Y$ set to 5 in receiver logic is enough.}

\jnote{check if we clearly state ``local fib propagation'' is not clearly defined in NDN}
}

\section{Conclusion}
\label{sec-conclusion}
Designing a congestion control mechanism
for information delivery to large numbers of receivers is
difficult, particularly with heterogeneous receiver bandwidths. It continues
to be a challenge also with Information Centric Networks.
Through emulation and an analytical model, we showed that heterogeneous receivers will get out-of-sync with existing receiver-driven, in-sequence congestion control mechanisms.
To overcome the resulting inefficiencies, we proposed \name.
\name allows receivers to request ``any-next'' packet, instead of the ``next in-sequence'' and thus delivers more packets on the first attempt.
For missing packets, other receivers can provide ``repair'' packets even if the in-network caches no longer hold the content.
Privacy and trust is maintained during the repair phase.

Our evaluations show that \name achieves efficiency and fairness on each path between the
information provider and receivers.
From a large scale simulation, \name can reduce aggregate network load (by up to $\sim$$46\%$) and transmission
completion times (by more than $50\%$) compared to existing congestion control mechanisms.
\name also reduces completion time by $\sim$$40\%$ while only increasing network load  by $\sim$$10\%$ compared to pgmcc, which aligns the sending rate to the slowest receiver.
Based on measurements on a prototype, \name outperforms ICP, getting 50\%
higher aggregate throughput and almost twice the throughput of pgmcc.
With the efficient repair of \name, streaming applications have
a much smaller \pt compared to the other mechanisms.


\section*{Acknowledgment}

The research leading to these results has received funding from the EU-JAPAN initiative by the EC Seventh Framework Programme (FP7/2007-2013) Grant Agreement No. 608518 (GreenICN), NICT under Contract No. 167, the US National Science Foundation under Grant No. CNS-1455815,  and the Volkswagen Foundation Project ``Simulation Science Center''. The views and conclusions contained herein are those of the authors and should not be interpreted as necessarily representing the official policies or endorsements, either expressed or implied, of the GreenICN project, the Simulation Science Center project, the European Commission, NSF, or NICT.

\bibliographystyle{IEEEtran}
\bibliography{ccnbib}

\end{document}